\documentclass[12pt,a4paper]{article}
\usepackage{graphicx}
\usepackage[english]{babel}
\usepackage{amsfonts}
\usepackage{amsmath}
\title{GENERAL FORMULAE OF INVARIANT FUNCTIONS OF THE GENERALIZED 
REACTION $\gamma N    \rightarrow    \gamma N$   IN   THE 
EFFECTIVE    LAGRANGIANS    METHOD}
\author{A. Yu. Loginov,\, V. N. Stibunov \\
{\it Nuclear Physics Institute} \\
{ \it at Tomsk Polytechnical University, Tomsk, Russia.}}
\date{}
\begin{document}
\maketitle
\begin{abstract}
The crossed channels of generalized  reaction $\gamma N \rightarrow \gamma N$
have been considered. 
The transformation coefficients from  the independent helicity
amplitudes    to    the   invariant  functions are calculated.
The  explicit  expressions  for invariant  functions have been 
obtained with the subject to the contribution of Born diagrams 
in $s$-,\, $u$-, and $t$-channel  and  six resonances  in $s$- 
and $u$-channel. It has been shown that the obtained invariant 
functions meet the requirements of crossing-symmetry. 
\end{abstract}

\section{}
\hspace*{\parindent}

Interaction   of  photons  with  nucleons is one of  the basic 
processes of elementary particle physics with initial energies 
up to 1 GeV. As an example we can point out the  pions photoproduction 
on  nucleons $\gamma N \rightarrow \pi N$,  bremsstrahlung off 
nucleon  $ N N \rightarrow N N \gamma$,  and,  finally,  Compton 
scattering  of  a photon on the nucleon $ \gamma N \rightarrow
\gamma N$.        This  paper deals with generalized  reaction  
$\gamma N \rightarrow \gamma N$  which includes  three crossed 
channels: 
Compton scattering of a photon on the nucleon $\gamma N \rightarrow \gamma N$  ($s$-channel), 
Compton scattering of a photon on the antinucleon $\gamma \bar{N} \rightarrow \gamma \bar{N}$  ($u$-channel), 
and annihilation of a nucleon-antinucleon pair into two photons 
$N \bar{N} \rightarrow \gamma \gamma $ ($t$-channel).

To describe the processes in the range from the threshold 
up to $\Delta (1232)$-isobar it is sufficient to take into 
account the contribution of the Born terms, vector mesons and  
the $\Delta (1232)$-isobar itself  \cite{olss1, olss2, blom}. 
However to describe the processes  in the broader energy area 
from  the  threshold  up  to 1 GeV  the calculation of higher 
nucleon resonances is indispensable. 
For  the  reaction  of  photoproduction  of  a pion on nucleon 
$\gamma N \rightarrow \pi N$ such calculation was carried  out 
in papers \cite{gar, drech}. 
In these  papers,  however, invariant  functions  of  reaction 
amplitude  carrying in themselves  the information on dynamics 
of the process were not calculated.  The availability  of  the 
explicit form of invariant  functions  significantly increases 
the  speed   of  numerical  simulation  of  processes  and  is 
indispensable  to  analytic calculate processes  amplitudes on 
nuclei. In this paper general analytical formulae for invariant 
functions of generalized reaction  $\gamma N \rightarrow \gamma N$                 
with  the  subject  to the contribution of Born terms in $s$-, 
$u$- and $t$-channel and six resonances in $s$- and $u$-channel 
have been calculated.

The article is organized as follows. 
In Section 2 the general expression for $P$- and $T$-invariant 
amplitude of reaction  $\gamma^{-}+\frac{1}{2}^{+}  \rightarrow
\gamma^{-}+\frac{1}{2}^{+}$  is given. 
In  Section  3   transformation  coefficients  from   helicity 
amplitudes of reaction   $\gamma^{-}+\frac{1}{2}^{+}  \rightarrow
\gamma^{-}+\frac{1}{2}^{+}$  to invariant functions are given. 
In Section 4 lagrangians  used  to  construct Feynman diagrams, 
in particular the problems of gauge-invariance  and connection 
with  the  fermion  field of spin $\frac{3}{2}$ are discussed.  
In   Section  5   the  properties  of   crossing-symmetry   of 
invariant functions and crossing  transformation of  amplitude 
from $s$-channel into a $u$-, and $t$-channel  are discussed. 
In  Section  6  the  Feynman  diagrams  which  have  been 
taken into account during amplitude calculation are shown.
Methods of calculation of the Born diagrams and  six resonances 
diagrams contribution in $s$-, $u$- and $t$-channel to invariant 
functions  are explained.
The Appendix lists the explicit  type  of lagrangians used and 
invariant functions obtained with their help.

\section{}
\hspace*{\parindent}

The  general  formulae  for  amplitude and invariant functions 
are  more convenient  to  obtain  considering  the reaction in 
$s$-channel $\gamma N  \rightarrow \gamma N$.    Transition to 
amplitudes of other channels can be then accomplished by means 
of crossing transformation of the amplitude of $s$-channel. 
To  construct  the  reaction  amplitude  of  $s$-channel it is 
necessary  to  define  the  number  of  independent  invariant
functions, which enter the expression for the amplitude.
It is equal to the number of independent helicity amplitudes 
of reaction $\gamma N \rightarrow \gamma N$  with the subject 
to $P$- and $T$-invariance.    The  total  number of helicity 
amplitudes of reaction $\gamma^{-}+\frac{1}{2}^{+} \rightarrow
\gamma^{-}+\frac{1}{2}^{+}$  is equal to                    
$2s_1 (2s_2+1) 2s_3 (2s_4+1) = 16$.                        
Between  these  amplitudes  there  are  ratios  following
from $P$-invariance of electromagnetic  interaction:
\begin{eqnarray}\label{e:1}
T(\lambda_3,\lambda_4;\;\lambda_1,\lambda_2)=
\eta(-1)^{(\lambda_1-\lambda_2)-(\lambda_3-\lambda_4)}
T(-\lambda_3,-\lambda_4;\;-\lambda_1,-\lambda_2),
\end{eqnarray}                                                                                         
where  $\eta=\eta_1 \eta_2 \eta_3 \eta_4 (-1)^{s_3+s_4-s_1-s_2}$;
$\eta_{i}, \, s_{i}$  are internal parities and spins of particles 
participating in the reaction, $\lambda_3, \lambda_4$ and  
$\lambda_1, \lambda_2$ are the helicities of the photon 
and   nucleon correspondently both in final and initial 
state.     It is easy to see, that the number of independent 
helicity amplitudes reduces to 8. 
The  further limitations on the number of independent helicity 
amplitudes  follow  from   $T$-invariance  of  electromagnetic 
interaction.  For  elastic  processes  $T$-invariance leads to 
the following ratios between helicity amplitudes:
\begin{eqnarray}\label{e:2}
T(\lambda_3,\lambda_4;\;\lambda_1,\lambda_2)=
(-1)^{(\lambda_1-\lambda_2)-(\lambda_3-\lambda_4)}
T(\lambda_1,\lambda_2;\;\lambda_3,\lambda_4)\, .
\end{eqnarray}
Thus the number of independent helicity amplitudes  of Compton 
scattering reduces to 6.   As  independent helicity amplitudes 
of Compton scattering we can select the following:
\begin{eqnarray}\label{e:3}
& & T(1,\frac{1}{2};1,\frac{1}{2}),\,
T(1, -\frac{1}{2};1,\frac{1}{2}),\,
T(-1,\frac{1}{2};1, \frac{1}{2}),\,
T(-1,-\frac{1}{2};1,\frac{1}{2}),\,
T(1,-\frac{1}{2};1,-\frac{1}{2}), \nonumber \\
& & T(-1,\frac{1}{2};1,-\frac{1}{2}) \,.
\end{eqnarray}                                                                                                                    
The  last  10  helicity  amplitudes  are  expressed  through 
independent helicity  amplitudes  \eqref{e:3}  according  to 
\eqref{e:1}, \eqref{e:2}:
\begin{eqnarray}
& & T(-1,-\frac{1}{2};-1,-\frac{1}{2}) =
T(1,\frac{1}{2};1,\frac{1}{2}), \,
T(-1,\frac{1}{2};-1,-\frac{1}{2}) =
- T(1,-\frac{1}{2};1,\frac{1}{2}), \nonumber  \\
& & T(1,\frac{1}{2};1,-\frac{1}{2}) =
-T(1,-\frac{1}{2};1,\frac{1}{2}), \,
T(-1,-\frac{1}{2};-1,\frac{1}{2}) =
T(1,-\frac{1}{2};1,\frac{1}{2}), \nonumber \\
& & T(1,-\frac{1}{2};-1,-\frac{1}{2}) =
T(-1,\frac{1}{2};1,\frac{1}{2}), \,
T(-1,-\frac{1}{2};1,-\frac{1}{2}) =
T(-1,\frac{1}{2};1,\frac{1}{2}), \nonumber  \\
& & T(1,\frac{1}{2};-1,\frac{1}{2}) =
T(-1,\frac{1}{2};1,\frac{1}{2}), \,
T(1,\frac{1}{2};-1,-\frac{1}{2}) =
- T(-1,-\frac{1}{2};1,\frac{1}{2}), \nonumber  \\
& & T(-1,\frac{1}{2};-1,\frac{1}{2}) =
T(1,-\frac{1}{2};1,-\frac{1}{2}), \,
T(1,-\frac{1}{2};-1,\frac{1}{2}) =
- T(-1,\frac{1}{2};1,-\frac{1}{2}) \, .
\end{eqnarray}

Consider  the  general  structure  of  amplitude  of   Compton 
scattering of a photon on nucleon $\gamma^{-}+\frac{1}{2}^{+}\rightarrow
\gamma^{-}+\frac{1}{2}^{+}$.                           
We  shall  designate  four-momenta of initial and final photons    
$k_1,\,k_2$  accordingly, and four-momenta of initial and final 
nucleons $p_1,\,p_2$ accordingly. We shall designate polarization 
four-vectors of initial and final photons $\epsilon_1,\,\epsilon_2$ 
and bispinors of initial and final nucleons -- $u_1,\,u_2$. 
To  find out the symmetry  of  the  amplitude  in  relation to 
spatial  reflection   and   time  reversion  it  is convenient 
to use symmetric and antisymmetric combination of four-momenta 
of photons: 
\begin{eqnarray}
K=\frac{1}{2}(k_1+k_2),\quad  Q=\frac{1}{2}(k_2-k_1).
\end{eqnarray}
Having added to $K$ and $Q$ a symmetric four-vector 
$P'= P-\frac{P\cdot K}{K^2} K, \;$            
where  $P=\frac{1}{2}(p_1+p_2)$ and pseudovector   
$N^{\mu} = i \epsilon^{\mu \nu \lambda \sigma} P'_{\nu} K_{\lambda} Q_{\sigma}$                              
we  shall  obtain  four  mutually  orthogonal  vectors with the 
help    of    which    it   is  convenient  to  describe   the 
gauge-invariant    structure    of  the  amplitude of reaction
$\gamma^{-}+\frac{1}{2}^{+}\rightarrow \gamma^{-}+\frac{1}{2}^{+}$. 
Invariant variables $s$, $t$ and $u$ are expressed 
through  four-vectors  $K,\, P,\, Q$  in  the  following  way:
\begin{eqnarray}
& & s=(P+K)^2,\quad t=4 Q^2, \quad u=(P-K)^2,
\quad  s+t+u=2M^2,
\end{eqnarray}
where $M$ -- is nucleon mass.
The general expression for the amplitude of 
reaction with arbitrary spins of particles, participating 
in it, looks like:
\begin{eqnarray}\label{e:8}
T(p_2,p_1;P)=\sum_{i} f_i(s,t)R^{i},
\end{eqnarray}
where   $f_i(s,t)$  are  the invariant   functions  depending 
on   four-momenta  of  initial  and  final  particles  through 
invariant variables  $s$   and  $t$  only,  $R^{i}$ -- invariant 
combinations, which should be  linearly composed of wave 
functions of all particles, participating in the  reaction. 
In case of reaction    $\gamma^{-}+\frac{1}{2}^{+}\rightarrow
\gamma^{-}+\frac{1}{2}^{+}$  the number of 
independent  invariant  functions   $f_{i}(s,t)$   with  the 
subject  to  $P$-  and  $T$-invariance are equal to 6. We can 
select  the  following  independent invariant combinations as:
\begin{eqnarray}\label{e:9}
& & R^{1}=\bar{u}(p_2)(P'\cdot \epsilon^{*}(k_2))
(P' \cdot \epsilon(k_1))P'^{-2} u(p_1),\;
R^{2}=\bar{u}(p_2)(P'\cdot \epsilon^{*}(k_2))
(P' \cdot \epsilon(k_1))\widehat{K} P'^{-2} u(p_1), \nonumber \\
& & R^{3}=\bar{u}(p_2)(N\cdot \epsilon^{*}(k_2))
(N \cdot \epsilon(k_1))N^{-2} u(p_1),\;
R^{4}=\bar{u}(p_2)(N\cdot \epsilon^{*}(k_2))
(N \cdot \epsilon(k_1))\widehat{K} N^{-2} u(p_1), \nonumber \\
& & R^{5}=\bar{u}(p_2)((P'\cdot \epsilon^{*}(k_2))
(N \cdot \epsilon(k_1)) -
(P'\cdot \epsilon(k_1))(N \cdot \epsilon^{*}(k_2))) \gamma_5
P'^{-2} K^{-2} u(p_1), \nonumber \\
& & R^{6}=\bar{u}(p_2)((P'\cdot \epsilon^{*}(k_2))
(N \cdot \epsilon(k_1)) +
(P'\cdot \epsilon(k_1))(N \cdot \epsilon^{*}(k_2)))
\gamma_5 \widehat{K} P'^{-2} K^{-2} u(p_1),
\end{eqnarray}
where $\widehat{K}=K_{\mu} \gamma^{\mu}$.

It  can  be  easily  seen  that  $R^{i}$  are   $P$-invariant. 
Let's   show   that   $R^{i}$   are   also  $C$-invariant. 
With  the  charge  conjugation  of  the  amplitude of reaction     
$\gamma^{-}+\frac{1}{2}^{+} \rightarrow \gamma^{-}+\frac{1}{2}^{+}$ 
the following permutations of wave functions and four-momenta 
of the particles are fulfilled \cite{nelipa}:
\begin{eqnarray}\label{e:010}
R^{i}_{k \, l}(\epsilon^{*}(k_2),k_2,p_2;\epsilon(k_1),k_1,p_1)
\rightarrow C_{k \, m} R^{i}_{n \, m}
(-\epsilon(k_1),-k_1,-p_1;-\epsilon^{*}(k_2),-k_2,-p_2)
C_{n \, l}.
\end{eqnarray}
Moreover there is a need to  transpose  $R^{i}_{k \, l}$ along 
the Dirac indexes $k$ and $l$ and to multiply from the left and 
from the right on the matrix of charge conjugation 
$ C = \gamma^2 \gamma^0 $. 
The resulting transformation of $R^{i}_{k \, l}$  may be presented 
in the following way:
\begin{eqnarray}\label{e:09}
R^{i}_{k \, l}(\epsilon^{*}(k_2),k_2,p_2;\epsilon(k_1),k_1,p_1)
\rightarrow C_{k \, m} R^{i}_{n \, m}
(-\epsilon(k_1),-k_1,-p_1;-\epsilon^{*}(k_2),-k_2,-p_2)
C_{n \, l}.
\end{eqnarray}
Using   the  properties  of   the  charge  conjugation  matrix
\cite{lifsh, itz} we can easily show that all the combinations 
\eqref{e:9} during the transformation \eqref{e:09} transfer into 
themselves, i.e. they are $C$-invariant.   Invariant variables 
$s,\, t,\, u$, and therefore invariant amplitudes $f_{i}(s,t)$ 
do not change during the transformation \eqref{e:010}. 
As  amplitude \eqref{e:8} is  $P$-  and   $C$-invariant,  it  is  
also $T$-invariant according to the $CPT$ theorem.

Invariant    combinations  $R^{i}$   are  at  the  same  time 
gauge-invariant, as all $R^{i}$ during the substitution 
$ \epsilon(k_1) \rightarrow k_1 $ or
$ \epsilon(k_2) \rightarrow k_2 $ turn to the zero. 
As   $k_1=K-Q,\;  k_2=K+Q $  we  have:
\begin{eqnarray} \label{e:0001}
& & P' \cdot k_1=
P'\cdot(K-Q)=P'\cdot K - P' \cdot Q = 0, \nonumber \\
& & P' \cdot k_2=
P'\cdot(K+Q)=P'\cdot K + P' \cdot Q = 0,
\end{eqnarray}
due to the orthogonality of four-vectors $P', K, Q$.       Scalar 
products  of  four-vectors  $N$  and  $k_1$, $k_2$  also equal to 
zero:
\begin{eqnarray} \label{e:0002}
& & N\cdot k_1 = i \epsilon^{\mu \nu \lambda \sigma} P'_{\nu}
K_{\lambda} Q_{\sigma} k_{1 \mu} =
i \epsilon^{\mu \nu \lambda \sigma} P'_{\nu}
K_{\lambda} Q_{\sigma} ( K_{\mu}-Q_{\mu}) = 0, \nonumber \\
& & N\cdot k_2 = i \epsilon^{\mu \nu \lambda \sigma} P'_{\nu}
K_{\lambda} Q_{\sigma} k_{2 \mu} =
i \epsilon^{\mu \nu \lambda \sigma} P'_{\nu}
K_{\lambda} Q_{\sigma} ( K_{\mu}+Q_{\mu}) = 0, 
\end{eqnarray}
as   the   products  of  tensors, one of which are absolutely 
antisymmetric   and   another   symmetric   in  two   indexes. 
It results to the fact that $R_{i}$ are gauge-invariant.

Finally let's write down the expression for $P$-, $C$-, $T$- and 
gauge-invariant helicity amplitude of the reaction 
$\gamma^{-}+\frac{1}{2}^{+}\rightarrow \gamma^{-}+\frac{1}{2}^{+}$:
\begin{eqnarray}\label{e:17}
& & T(\lambda_3, \lambda_4,\; \lambda_1, \lambda_2)=
f_1(s,t) \bar{u}(p_2,\lambda_4)(P'\cdot \epsilon^{*}(k_2, \lambda_3))
(P' \cdot \epsilon(k_1, \lambda_1)) P'^{-2} u(p_1, \lambda_2) + \nonumber \\
& & + f_2(s,t)\bar{u}(p_2,\lambda_4)(P'\cdot \epsilon^{*}(k_2,\lambda_3))
(P' \cdot \epsilon(k_1,\lambda_1) )\widehat{K} P'^{-2}
u(p_1,\lambda_2) + \nonumber \\
& & + f_3(s,t)\bar{u}(p_2,\lambda_4)(N\cdot \epsilon^{*}(k_2,\lambda_3))
(N \cdot \epsilon(k_1,\lambda_1))N^{-2} u(p_1, \lambda_2) + \nonumber \\
& & + f_4(s,t)\bar{u}(p_2,\lambda_4)(N\cdot \epsilon^{*}(k_2,\lambda_3))
(N \cdot \epsilon(k_1,\lambda_1))\widehat{K} N^{-2} u(p_1,\lambda_2) + \nonumber \\
& & + f_5(s,t)\bar{u}(p_2,\lambda_4)
((P'\cdot \epsilon^{*}(k_2,\lambda_3))
(N \cdot \epsilon(k_1,\lambda_1)) - \nonumber \\
& & - (P'\cdot \epsilon(k_1,\lambda_1))(N \cdot \epsilon^{*}(k_2,\lambda_3))) \gamma_5
P'^{-2} K^{-2} u(p_1,\lambda_2) + \nonumber \\
& & + f_6(s,t)\bar{u}(p_2,\lambda_4)
((P'\cdot \epsilon^{*}(k_2,\lambda_3))
(N \cdot \epsilon(k_1,\lambda_1)) + \nonumber \\
& & + (P'\cdot \epsilon(k_1,\lambda_1))(N \cdot \epsilon^{*}(k_2,\lambda_3)))
\gamma_5 \widehat{K} P'^{-2} K^{-2} u(p_1,\lambda_2).
\end{eqnarray}
In  definition  of  helicity  bispinors  of  the  nucleons 
the convention \cite{jacob} about phase of second particle
is taken into account.

\section{}
\hspace*{\parindent}

Let's   get  the  matrix  of  the  transformation  between the 
six independent helicity amplitudes  \eqref{e:3} and six invariant 
function $f_{i}(s,t)$.         Having numbered the independent 
helicity amplitudes \eqref{e:3} from  1 to 6 we may write them 
down in the explicit form. In order to do this there is a need 
to  put  into  the general expression of the helicity amplitude 
\eqref{e:17}  the explicit expressions for four-vectors $K, Q, P'$ 
and $N$  and  also  for the helicity bispirons of the nucleons 
$u(p_1,\lambda_2),\, u(p_4,\lambda_4)$, taking into consideration 
the convention \cite{jacob}, and four-vectors of photon polarization  
$\epsilon(k_1,\lambda_1),\,\epsilon(k_2,\lambda_3)$. 
As a result we will have a linear heterogeneous equation system 
of the following type:  
\begin{eqnarray}\label{e:18}
T_{i}(s,t) = \sum_{j=1}^{6} A_{i\,j} f_{j}(s,t),
\end{eqnarray}
where $ T_{i}(s,t) $ are independent helicity amplitudes \eqref{e:3}.

With   the  help  of  the  package of  symbolic calculations 
Mathematica  it  is  possible to show that  the  determinant 
of  the  system  matrix \eqref{e:18}  is  not  equal to zero. 
This  may  serve  as  a proof for the linear independence of 
invariant combinations \eqref{e:9}. Using the package Mathematica 
we may solve the linear heterogeneous  system \eqref{e:18} 
with regard to invariant functions  $f_{i}(s,t)$:
\begin{eqnarray}
f_{i}(s,t) = \sum_{j=1}^{6} u_{i\,j} T_{j}(s,t).
\end{eqnarray}
In the matrix $\|u_{i\,j}\|$ 8 elements out of 36 
are equal to zero and the remaining 28 nonzero elements may 
be divided into 4 groups:
\begin{eqnarray}\label{e:019}
& & u_{1\,1}=-\frac{1}{2} u_{1\,3}= u_{1\,5}=-u_{3\,1}=
-\frac{1}{2} u_{3\,3} =
-u_{3\,5}= M u_{6\,1} = -M u_{6\,5} =
-\frac{M^2\,\sec (\frac{\theta }{2})}{s - M^2}, \nonumber \\
& & u_{1\,2} = -2 u_{1\,4} = 2 u_{1\,6} = -u_{3\,2} =
-2 u_{3\,4} = 2 u_{3\,6} = -2 u_{5\,4} = -2 u_{5\,6} =
\frac{2\,M\,{\sqrt{s}}\,\csc (\frac{\theta }{2})}
{s-M^2}, \nonumber \\
& & u_{2\,1} = -\frac{1}{2} u_{2\,3} = u_{2\,5}=-u_{4\,1}=
-\frac{1}{2} u_{4\,3} = -u_{4\,5} =
\frac{M\,\left( M^2 + s \right) \,\sec (\frac{\theta }{2})}
{{\left( M^2 - s \right) }^2}, \nonumber \\
& & u_{2\,2}=-2 u_{2\,4}=2 u_{2\,6}=-u_{4\,2}=-2 u_{4\,4}=
2 u_{4\,6} =  -\frac{4\,M^2\,{\sqrt{s}}\,\csc (\frac{\theta }{2})}
{{\left( M^2 - s \right) }^2}, \nonumber \\
& & u_{5\,1}=u_{5\,2}=u_{5\,3}=u_{5\,5}=u_{6\,2}=u_{6\,3}=
u_{6\,4}=u_{6\,6}=0 .
\end{eqnarray}
In the expression \eqref{e:019} $\theta$ is the scattering 
angle of photon in the center of mass system of the 
$s$-channel:
\begin{eqnarray}
\cos(\theta)=1 + \frac{2\,s\,t}{{\left( M^2 - s \right) }^2}.
\end{eqnarray}

\section{}
\hspace*{\parindent}

In   the   present   work   we  took  into  consideration  the 
contribution into the amplitude  of  Born  terms in $s$-, $u$- 
and $t$-channel and the contribution of six resonances 
$P_{33}(1232),\\ P_{11}(1430),\, S_{11}(1500),\,
D_{13}(1505),\, S_{31}(1620),\, D_{33} (1700)$ in $s$- and $u$-channel. 
At the same time we face the question of choice of corresponding 
lagrangians of interaction, that correspond to the vertices of
Feynman diagrams. 
In this work while choosing the lagrangians of interaction  no 
simplifying  assumptions  were  made,  i.e.  langrangians   of 
interaction  was  written in the most general form, compatible 
with the requirements of hermiticity, $P$-, $T$- and $C$-invariance. 
As for the gauge-invariance we may say that
the lagrangians of interaction with $\gamma$-quantum, which 
the electromagnetic field enters by means of gauge-invariant 
tensor $F_{\mu \nu} = -i (k_{\mu}\epsilon_{\nu}- 
k_{\nu}\epsilon_{\mu})$ are also gradient invariant, regardless 
of whether they are on mass shell or not, because during the 
substitution  $\epsilon \rightarrow k$ tensor $F_{\mu \nu}$ 
identically turns to zero.  Almost all the lagrangians, used 
in this work, refer to this type, except the Dirac's part of 
the  interaction  lagrangian    of   photon   with   nucleon 
\eqref{e:90}. This last 
lagrangian is gauge-invariant only when two nucleons are on 
mass shell, which is the consequence of the  Dirac equation 
for free nucleon. If one or both nucleons are not on the mass 
shell, the vertex corresponding to the Dirac's part of interaction 
lagrangian of  photon  with  nucleon \eqref{e:90} formally 
won't be gauge-invariant. However it can be easily shown that 
the sum of two nucleon Born diagrams in $s$- and $u$-channels 
will be gauge-invariant the same as for Compton scattering on 
the electron, thus all the observables of reactions will be 
gauge-invariant. As the lagrangian $\pi N$ interaction in the 
present work we have used pseudoscalar variant. We have to 
mention that in case of Compton scattering in the Born diagram 
of $t$-channel, which corresponds to the interaction of pion 
with nucleon, both nucleons are on mass shell, that's why 
pseudovector variant $\pi N$  interaction in this case is 
equivalent to pseudoscalar one. The difference between the 
pseudoscalar and pseudovector coupling  occurs  in case of 
Compton scattering only when observing diagrams with loops, 
in which one or both nucleons aren't on the mass shell. At the 
same time the usage of pseudovector   coupling  leads to the 
extra    degrees   of   momentum   in   the   numerator   of 
integrand, which worsens the degree of  integral convergence. 
That's why from the mathematical point of view the usage of 
pseudoscalar connection in loop diagrams is more preferable. 
In contrast to this, in Born diagrams of photoproduction of 
pion on the nucleon, one of the nucleons of $\pi N N$ vertex 
always is beyond the mass shell, which leads to the nonequivalence 
of pseudoscalar and pseudovector coupling already on the level 
of Born diagrams. 

Let's  consider  lagrangian,  which  corresponds  to  the 
possible  contact four-particle interaction of two photons 
and two nucleons. This lagrangian come from Pauli's part of 
interaction lagrangian of  photon  with  nucleon \eqref{e:90}
with aid of  minimal substitution 
$\partial_{\mu} \rightarrow \partial_{\mu} - i e A_{\mu}$
and  should have the following form: 
\begin{equation}\label{e:20}
\mathcal{L}_{cont}=\frac{e^2}{2 M} A^{\mu}
\overline{N}\sigma_{\mu \nu} N A^{\nu}.
\end{equation}
We can show that the vertex, corresponding to the lagrangian 
\eqref{e:20} during the gauge replacement 
$\epsilon^{\mu} \rightarrow k^{\mu}$, in case when nucleons 
are on mass shell, doesn't turn identically to zero but is 
the value, proportional to the multiplier  $(s-M^2)$.
Thus, in relativistic theory of perturbation the lagrangian 
\eqref{e:20} is not gauge-invariant. It is important to note, 
however, that there are also contact terms in the recording 
of three-momentum Compton scattering amplitude in the center 
of mass system, proportional to 
$\mbox{\boldmath$\epsilon$}_{1}  \cdot \mbox{\boldmath$\epsilon$}_{2}$, 
or
$\mbox{\boldmath$\epsilon$}_{1} \times \mbox{\boldmath$\epsilon$}_{2}$. 
But they  occur as three-dimension reduction of non-contact 
relativistic diagrams.

Three of six resonances, which contribution is taken 
into account   in   the  $s$- and $u$-channel,  have 
a spin equal to $\frac{3}{2}$. The Rarita-Schwinger 
propagator, corresponding to these resonances, describes the 
transfer  between  the  $\frac{3}{2}$  spins  states only on 
the  mass shell  $P^2={M^{*}}^{2}$ and does not possess this 
property when the particle with  spin $\frac{3}{2}$ is not 
located on the mass shell. If such propagator is used when 
the particle with  $\frac{3}{2}$ spin  does not lie on the 
mass shell it is necessarily to provide interaction lagrangian 
with additional condition, ensuring the connection only with
the $\frac{3}{2}$ spin field. Let us consider the interaction 
of the $\frac{3}{2}$ spin particle with the tensor of electromagnetic 
field  $F^{\mu \nu}$. In this case the interaction lagrangian 
has  the following expression:
\begin{equation}
\mathcal{L}_{int}=\overline{N}^{\ast \mu}
O_{\mu \nu \sigma}N F^{\nu \sigma} + h. c.,
\end{equation}
where $N^{\ast \mu}$ is the field of $\frac{3}{2}$ spin, 
$N$ --  a nucleon field. 
In order to ensure the connection with the $\frac{3}{2}$ spin 
only, vertex matrix $O_{\mu \nu \sigma}$ must meet the following 
requirements \cite{gar, peccei}:
\begin{equation}\label{e:22}
\gamma^{\mu}O_{\mu \nu \sigma}=0.
\end{equation}  
Acquiring this condition is possible under the following 
circumstances. Let the vertex matrix $\Gamma_{\mu \nu \sigma}$ 
possess all required  properties of relativistic invariance and 
$C$-, $P$-, $T$-invariance on the mass shell.       Now let us 
write \cite{peccei} the new vertex matrix: 
\begin{equation}
O_{\mu \nu \sigma}=\Gamma_{\mu \nu \sigma}-
\frac{1}{4}\gamma_{\mu}\gamma^{\eta}
\Gamma_{\eta \nu \sigma}.
\end{equation}
Matrix $O_{\mu \nu \sigma}$ will obviously meet the requirement 
\eqref{e:22}, and  when  located  on the mass shell it becomes 
equal to matrix $\Gamma_{\mu \nu \sigma}$, because the 
Rarita-Schwinger spinors $N^{\ast \mu}$ 
meet the requirement of  $ \gamma_{\mu} N^{\ast \mu} = 0$. 
All interaction lagrangians of the $\frac{3}{2}$ spins particles 
used in this paper meet requirement \eqref{e:22}. Lagrangians, 
used  for  the  diagrams  construction,     and  the  explicit 
form of the propagators of particles with spin $\frac{1}{2}$ and 
$\frac{3}{2}$ is shown in the Appendix.

\section{}
\hspace*{\parindent}

If  $\gamma     N     \rightarrow   \gamma  N$  is regarded as 
the  generalized  reaction, which takes place in three crossed 
channels  $s$-, $u$- and $t$-, it is possible to established 
crossing-symmetry properties of the invariant amplitudes $f_{i}$. 
The  crossing-symmetry of the arbitrary  generalized reaction 
is  possible  if  four  particles  include  two identical ones 
(the particles, which relate to one isomultiplet, are 
regarded as identical, for example, nucleons and pions).
In our case there are two pairs of identical particles -- 
two photons and two nucleons.   The crossing of any of 
these particle pairs results in transformation of $s$-channel 
into $u$-channel.   In this case, due to generalized Pauli's 
principle, the received amplitude must be identical to the 
initial  one  during  the  crossing  of  two  photons
and differ only  in  sign during the crossing of two 
nucleons.  
Examining of two photons crossing is less complicated. 
To carry out the crossing of two photons in initial 
amplitude  \eqref{e:17} the following changes would 
be necessary \cite{nelipa}:
\begin{eqnarray}\label{e:50}
& & k_1 \rightarrow - k_2,\quad k_2 \rightarrow -k_1, \quad
K \rightarrow -K, \quad P \rightarrow P,  \quad P' \rightarrow
P', \quad Q \rightarrow Q, \quad N \rightarrow -N, \nonumber \\
& & s \rightarrow u, \quad u \rightarrow s, \quad t \rightarrow t,\quad
\epsilon(k_1) \rightarrow - \epsilon^{*}(k_2), \quad \epsilon(k_2) \rightarrow - \epsilon^{*}(k_1).
\end{eqnarray}
Here the invariant combinations $R^{i}$ \eqref{e:9} are converted 
as  follows: 
\begin{eqnarray}\label{e:51}
R^{1} \rightarrow R^{1}, \quad R^{2} \rightarrow -R^{2},\quad
R^{3} \rightarrow R^{3}, \quad R^{4} \rightarrow -R^{4},\quad
R^{5} \rightarrow R^{5}, \quad R^{6} \rightarrow  R^{6}.
\end{eqnarray}  
In   view  of   generalized   Pauli's   principle  during  the 
permutation of two identical photons, the amplitude \eqref{e:17} 
must not change its sign, the invariant functions $f_1 - f_6$ 
are converted as follows: 
\begin{eqnarray}\label{e:52}
& & f_{1}(s,t) \rightarrow f_{1}(u,t), \quad f_{2}(s,t) \rightarrow -f_{2}(u,t),\quad
f_{3}(s,t) \rightarrow f_{3}(u,t), \quad f_{4}(s,t) \rightarrow -f_{4}(u,t),\nonumber\\
& & f_{5}(s,t) \rightarrow f_{5}(u,t), \quad f_{6}(s,t) \rightarrow  f_{6}(u,t).
\end{eqnarray}
The definition of the invariant functions symmetry 
properties $f_1 - f_6$ with transposition $ s \leftrightarrow u $ 
allows accomplishing the crossing of  identical nucleons 
instead  of  the  crossing  of  identical photons on the 
same base.
In  is case instead of \eqref{e:50} we will have:
\begin{eqnarray}
& & p_1 \rightarrow - p_2,\quad p_2 \rightarrow -p_1, \quad
K \rightarrow K, \quad P \rightarrow -P, \quad P' \rightarrow
- P', \quad Q \rightarrow Q, \quad N \rightarrow -N, \nonumber \\
& & s \rightarrow u, \quad u \rightarrow s, \quad t \rightarrow t, \quad
u(p_1) \rightarrow u(-p_2) \equiv v(p_2), \quad
u(p_2) \rightarrow u(-p_1) \equiv v(p_1), \\
& & \overline{u}_{i}(p_2) M_{i\,j}(p_2,p_1) u_{j}(p_1)
\rightarrow - \overline{v}_{i}(p_1) M_{i\,j}(-p_1,-p_2)
v_{j}(p_2) = \overline{u}_{i}(p_2) \widetilde{M}_{i\,j}(-p_1,-p_2)
u_{j}(p_1), \nonumber
\end{eqnarray}
where   the  last  line  has  the  equalities  of    $ v(p) =
C \widetilde{\overline{u}}(p),\, \overline{v}(p) = - \widetilde{u}(p) C $
and the matrix of charge conjugation properties \cite{lifsh, itz}. 
It is clear that under this condition the invariant spin combinations 
$R^{i}$  
\eqref{e:9} are converted similarly as in \eqref{e:51}, and we 
again may state that the invariant functions $f_i$ have the 
crossing-symmetry properties as in \eqref{e:52}.
 
Along with the invariant functions calculation we considered 
the reaction in $s$-channel  $\gamma N \rightarrow \gamma N$. 
Accomplishing the crossing transformation provided transition 
of the reaction amplitude in $s$-channel \eqref{e:17} 
to the  reaction amplitude  in  $u$- and $t$-channel. 
The  transition  of  the  amplitude \eqref{e:17} in  $s$-channel 
$\gamma(k_1) N(p_1) \rightarrow \gamma(k_2) N(p_2)$  into  the 
amplitude of reaction in $u$-channel 
$\gamma(k_1) \bar{N}(\overline{p}_1) \rightarrow \gamma(k_2)
\bar{N}(\overline{p}_2)$,  would   require   the    following 
replacements: 
\begin{eqnarray}\label{e:80}
p_1 \rightarrow - \overline{p}_2, \quad
p_2 \rightarrow - \overline{p}_1, \quad
u(p_1) \rightarrow u(-\overline{p}_2) \equiv v(\overline{p}_2), \quad
\overline{u}(p_2) \rightarrow \overline{u}(-\overline{p}_1) \equiv
\overline{v}(\overline{p}_1).
\end{eqnarray}
In  this  case  four-vectors  $K$, $Q$,  remain constant, 
and  four-vectors $P, P', N$  as  well  as the  invariant 
variables  $s,  t,  u$   are   determined   as   follows:
\begin{eqnarray}
& & P=-\frac{1}{2} (\overline{p}_1 + \overline{p}_2), \quad
P'= P-\frac{P\cdot K}{K^2} K, \quad
N^{\mu} = i \epsilon^{\mu \nu \lambda \sigma} P'_{\nu}
K_{\lambda} Q_{\sigma}, \nonumber \\
& & s=(k_1-\overline{p}_2)^2,\quad t=(k_1-k_2)^2, \quad
u=(\overline{p}_1+k_1)^2.
\end{eqnarray} 
The transition of the $s$-channel reaction amplitude  into 
the $t$-channel reaction amplitude
$N(p) \bar{N}(\overline{p}) \rightarrow \gamma(k) \gamma(k')$,
requires the following replacements in \eqref{e:17}:
\begin{eqnarray}\label{e:81}
& & p_1 \rightarrow p, \quad
p_2 \rightarrow - \overline{p}, \quad
k_1 \rightarrow -k, \quad
k_2 \rightarrow k', \quad
u(p_1) \rightarrow u(p), \nonumber \\
& & \overline{u}(p_2) \rightarrow \overline{u}(-\overline{p})
\equiv \overline{v}(\overline{p}),\quad
\epsilon(k_1) \rightarrow \epsilon(-k) = -\epsilon^{*}(k),
\quad \epsilon^{*}(k_2) \rightarrow \epsilon^{*}(k').
\end{eqnarray}  
Four-vectors  $K,  Q,  P,  P',  N$   and   the   invariant 
variables $s, t, u$ in this case are determined as follows: 
\begin{eqnarray}
& & K=\frac{1}{2}(k'-k),\quad Q=\frac{1}{2}(k+k'),\quad
P=\frac{1}{2} (p - \overline{p}), \quad
P'= P-\frac{P\cdot K}{K^2} K, \nonumber \\
& & N^{\mu} = i \epsilon^{\mu \nu \lambda \sigma} P'_{\nu}
K_{\lambda} Q_{\sigma},\quad s=(p-k)^2,\quad
t=(k + k')^2, \quad u=(\overline{p} - k)^2.
\end{eqnarray}

\section{}
\hspace*{\parindent}

When    calculating    invariant    functions   $f_1 - f_6$           
the amplitude of Compton scattring reaction was recorded in 
accordance  with  Feynman  rules  for  diagrams in figure 1.
\begin{figure}
\begin{center}
\includegraphics[width=12cm]{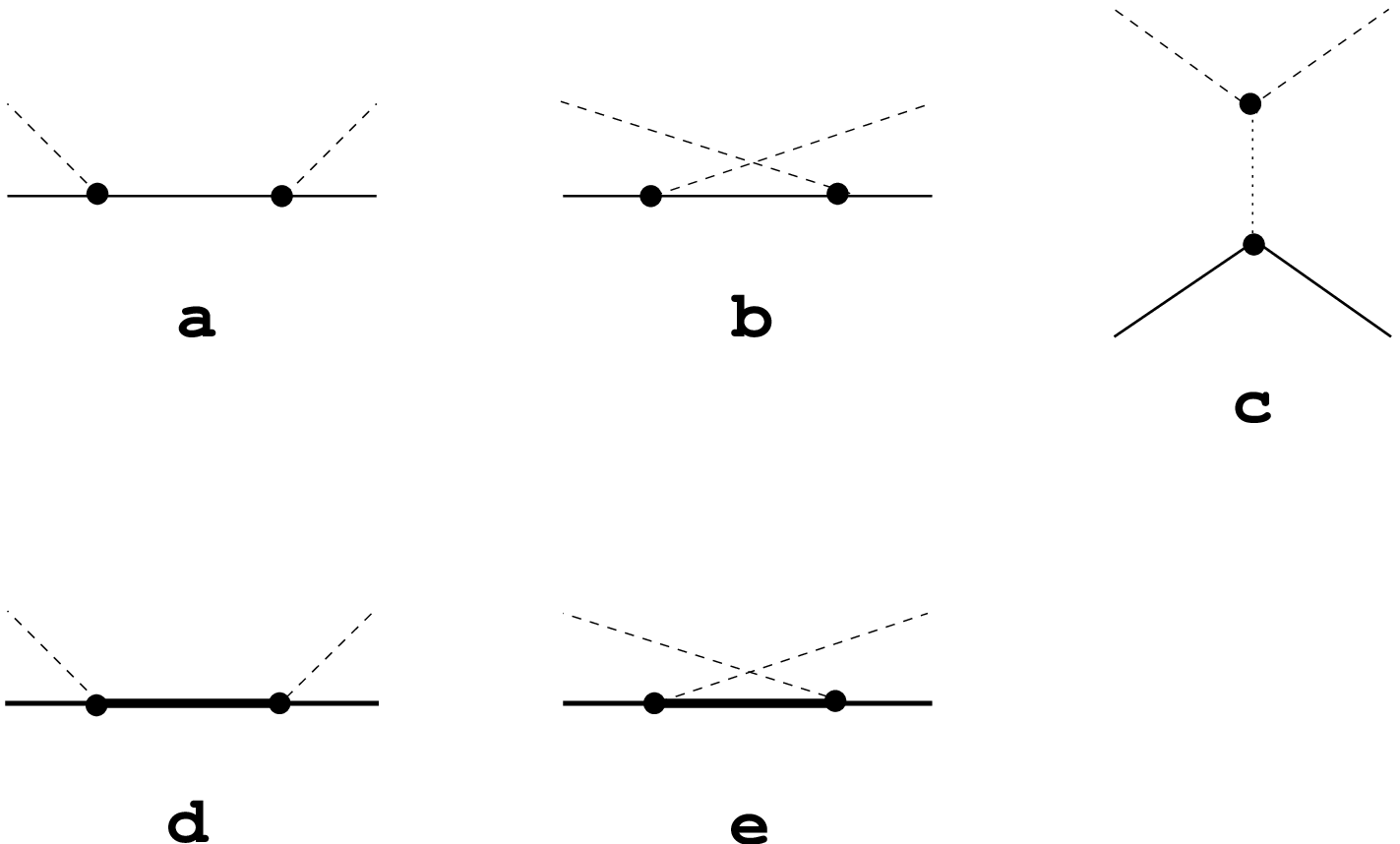}
\end{center}
\begin{center}
\parbox[l]{13cm}{
\begin{center}
{\large{\textbf {Fig.1}}} 
\end{center}
Diagrams   are  considered  on  calculation  of  the invariant 
functions $f_1 - f_6$.   Diagrams    \textbf{a,  b,  c}    are 
corresponding  to  Born  terms  in  $s$-,  $u$-  and $t$-channel, 
diagrams \textbf{d, e} are corresponding to resonances terms in 
$s$- and $u$-channel.}
\end{center}
\end{figure} 

Vertices  of  diagrams  correspond  to  lagrangians are  cited 
in the Appendix. Propagators of $\frac{1}{2}$ and $\frac{3}{2}$ 
spin resonances, used when recording the amplitude,  are cited 
in \eqref{e:66}, \eqref{e:67}.

Using direct expressions for helicity bispinors and four-vectors 
of  photon  polarization  \cite{haber}  it  is possible to get 
expressions for six independent helicity amplitudes \eqref{e:3} 
of Compton scattering with the help of symbolic calculations 
package Mathematica. 
Then using the calculated transformation matrix \eqref{e:019} of 
independent  helicity  amplitudes  \eqref{e:3}  to  invariant 
functions $f_1 - f_6$ it is possible to get explicit expressions 
for the latter. 
Calculated contributions of Born diagrams in $s$-, $u$- and 
$t$-channel  and  contributions  of  resonances  diagrams 
$P_{33}(1232),\, P_{11}(1430),\,
S_{11}(1500),\,  S_{31}(1620), \, D_{33}(1700),\, D_{13}(1505)$       
in $s$- and $u$-channel into invariant  functions  of Compton 
scattering are sited in the Appendix. 
Cases concerning Compton scattering on proton and neutron are 
considered.
Invariant variables $s, \, t$ are used as independent arguments 
of invariant functions  $f_1 - f_6$.  All the given expressions 
of contributions in invariant functions  meet the requirements 
for crossing symmetry \eqref{e:52}. 
Invariant functions $f_1 - f_6$  of Compton  scattering  after 
crossing   transformation  \eqref{e:80}, \eqref{e:81}  may  be 
used for the construction of reaction amplitude in $u$-channel 
and $t$-channel correspondingly.

This work was supported by the Russian Basic Research
Foundation (grants N   01-02-17276 and 03-02-06280) and
the RF Ministry of Education (grant N  E02-3.3-216).
\vspace{2 cm}

\appendix{\LARGE{\textsc{Appendix}}}
\vspace{1 cm}\\

Below  there  is  a  explicit  structure  of    interaction 
lagrangians,  used  in the work.  The following indications 
are accepted: $N^{\ast \mu}$ is  $\frac{3}{2}$  spin  field, 
which  corresponds  to  the  given  resonance,   $N^{\ast}$          
is $\frac{1}{2}$ spin field, which corresponds to the given 
resonance, $N$  is nucleon field, $\pi_{i}$ is pionic field 
with isotopic index $i$,  $\eta$  is     $\eta$-meson field.               
$A^{\mu}$ is four-potential of electromagnetic field,                               
$F^{\mu \nu}=\partial^{\mu} A^{\nu} - \partial^{\nu} A^{\mu}$
is gauge-invariant tensor of electromagnetic field.
Mass  indications:  $M$     is   nucleon   mass, $\mu$                 
is   pion   mass,  $\mu_{\eta}$ is $\eta$-meson  mass. 
Dirac   matrixes   $\gamma_{\mu}$,\, $\gamma_5$   and 
antisymmetric matrix tensor $\sigma_{\mu \nu}$ are determined 
as in \cite{bjork}. 
The   following   indications   were   used  in  isotopic part 
of lagrangians:  $\tau_{i}$ is isotopic Pauli matrix,  $S_{i}$
is isotopic transition matrix   $\frac{3}{2} \rightarrow \frac{1}{2}$ 
\cite{eric}.
 
Lagrangian  of  photon  interaction  with  nucleon:
\begin{eqnarray}\label{e:90}
& &  \mathcal{L}_{N N \gamma}=
\overline{N}\frac{1}{2}\left(F_{1}^{S}+F_{1}^{V}\tau_{3}\right)
\widehat{A} N  -
\frac{F^{\alpha \beta }}{4M}\overline{N}\sigma _{\alpha
\beta }\frac{1}{2}\left( F_{2}^{S}+F_{2}^{V}\tau _{3}\right) N,
\end{eqnarray}
where   $F_{i}^{S}   =   F_{i}^{p} + F_{i}^{n},\, F_{i}^{V} = F_{i}^{p}
- F_{i}^{n}$  are  isoscalar  and isovector  form  factors  of 
nucleon, $\widehat{A}=A_{\alpha}\gamma^{\alpha}$.

Lagrangians  of  interaction,  used  in  the  calculations  of 
Born  diagrams  in  $t$-channel  of  Compton  scattering:
\begin{eqnarray}
\mathcal{L}_{\pi \gamma \gamma }=\frac{1}{4}F^{\pi }\epsilon ^{\mu \nu
\alpha \beta }F_{\mu \nu }F_{\alpha \beta }\pi_{0}, \label{e:001} \\
\mathcal{L}_{\pi N N}=\frac{2 i M f}{\mu}\overline{N}
\gamma_{5}\tau_{i} N \pi _{i}, \label{e:002} \\
\mathcal{L}_{\eta \gamma \gamma }=\frac{1}{4}F^{\eta }
\epsilon ^{\mu \nu \alpha \beta }F_{\mu \nu }
F_{\alpha \beta }\eta, \label{e:003} \\
\mathcal{L}_{\eta N N}=\frac{2 i M f_{\eta}}{\mu_{\eta}}\overline{N}
\gamma_{5} N \eta,\label{e:004}
\end{eqnarray}
where  $F^{\pi },\, F^{\eta},\, f,\, f_{\eta}$  are corresponding 
coupling constants. Lagrangians \eqref{e:002}, \eqref{e:004} correspond 
to pseudoscalar  meson-nucleon coupling, lagrangians  \eqref{e:001}, 
\eqref{e:003} are obviously gauge-invariant.

Lagrangian of   $NS_{11}\gamma$   interaction:
\begin{equation}\label{e:60}
\mathcal{L}_{NS_{11}\gamma }=-\frac{F^{\alpha \beta }}{4M}\left[ \overline{N}%
\sigma _{\alpha \beta }\gamma _{5}\left( G_{{NS_{11}\gamma }}^{S}+\tau
_{3}G_{{NS_{11}\gamma }}^{V}\right) N^{\ast }-\overline{N}^{\ast }\sigma
_{\alpha \beta }\gamma _{5}\left( G_{{NS_{11}\gamma }}^{S}+\tau
_{3}G_{{NS_{11}\gamma }}^{V}\right) N\right],
\end{equation} 
where $G_{{NS_{11}\gamma }}^{S} = G_{{NS_{11}\gamma }}^{p} + \; G_{{NS_{11}\gamma }}^{n},\,
G_{{NS_{11}\gamma }}^{V} = G_{{NS_{11}\gamma }}^{p} - \; G_{{NS_{11}\gamma }}^{n}$ 
are isoscalar and isovector form factors of $N S_{11}\gamma$
interaction.

Lagrangian of  $NS_{31}\gamma$ interaction:
\begin{equation}\label{e:61}
\mathcal{L}_{NS_{31}\gamma }=-\frac{G_{{NS_{31}\gamma}}}{2M}F^{\alpha \beta }%
\left[ \overline{N}\sigma _{\alpha \beta }\gamma _{5}S_{3}N^{\ast }-%
\overline{N}^{\ast }\sigma _{\alpha \beta }\gamma _{5}S_{3}^{\dag }N\right],
\end{equation}
where $G_{{NS_{31}\gamma}}$ is isovector form factor of $NS_{31}\gamma$ 
interatcion. 
In this case isoscalar  form  factor  is  absent,  because 
absorption or emission of only isovector  $\gamma$-quantum 
is  possible  in  $NS_{31}\gamma$  vertex.   The  same  may 
be said about other resonances  with  $\frac{3}{2}$ isospin.
 
Lagrangian of  $NP_{11}\gamma$ interaction:
\begin{equation}\label{e:62}
\mathcal{L}_{NP_{11}\gamma }=-\frac{F^{\alpha \beta }}{4M}\left[ \overline{N}%
\sigma _{\alpha \beta }\left( G_{{NP_{11}\gamma }}^{S}+\tau
_{3}G_{{NP_{11}\gamma }}^{V}\right) N^{\ast }+\overline{N}^{\ast }\sigma
_{\alpha \beta }\left( G_{{NP_{11}\gamma }}^{S}+\tau _{3}G_{{NP_{11}\gamma
}}^{V}\right) N\right],
\end{equation}
where $G_{{NP_{11}\gamma }}^{S} = G_{{NP_{11}\gamma }}^{p} + \; G_{{NP_{11}\gamma }}^{n},\,
G_{{NP_{11}\gamma }}^{V} = G_{{NP_{11}\gamma }}^{p} - \; G_{{NP_{11}\gamma }}^{n}$ 
are isoscalar and isovector  form  factors  of $NP_{11}\gamma$ 
interaction.

Lagrangian of $NP_{33}\gamma$ interaction:
\begin{eqnarray}\label{e:63}
& & \mathcal{L}_{NP_{33}\gamma }=\frac{G_{1NP_{33}\gamma }}{4M}\left\{ \overline{%
N}^{\ast \lambda }\left[ \left( g_{\lambda \alpha }\gamma _{\beta
}-g_{\lambda \beta }\gamma _{\alpha }\right) +\frac{i}{2}\gamma _{\lambda
}\sigma _{\alpha \beta }\right] \gamma _{5}S_{3}^{\dag }N+h.c.\right\}
F^{\alpha \beta }- \nonumber \\
& & - \frac{G_{2NP_{33}\gamma }}{8M^{2}}\left\{ \overline{N}^{\ast \lambda }\left[
i\left( g_{\lambda \beta } \overleftrightarrow{\partial_{\alpha}}-g_{\lambda \alpha }%
\overleftrightarrow{\partial_{\beta }}\right) - \frac{i}{4}\gamma_{\lambda }\left(
\gamma _{\beta }\overleftrightarrow{\partial_{\alpha }}-\gamma _{\alpha }%
\overleftrightarrow{\partial_{\beta }}\right) \right] \gamma _{5}S_{3}^{\dag
}N+h.c.\right\} F^{\alpha \beta }+ \nonumber \\
& & +\frac{G_{3NP_{33}\gamma }}{4M^{2}}\left\{ \left[ \overline{N}^{\ast \lambda
}\left( -g_{\lambda \beta }+\frac{1}{4}\gamma _{\lambda }\gamma _{\beta
}\right) i\gamma _{5}S_{3}^{\dag }N+h.c.\right] \partial _{\alpha }F^{\alpha
\beta }+ \right. \nonumber\\
& & \left. +\left[ \overline{N}^{\ast \lambda }\left( g_{\lambda \alpha }-\frac{1%
}{4}\gamma _{\lambda }\gamma _{\alpha }\right) i\gamma _{5}S_{3}^{\dag
}N+h.c.\right] \partial _{\beta }F^{\alpha \beta }\right\},
\end{eqnarray}
where $G_{1NP_{33}\gamma},\, G_{2NP_{33}\gamma},\, G_{3NP_{33}\gamma}$ 
are  isovector  form  factors of  $NP_{33}\gamma$ interaction. 
Symbol $\overleftrightarrow{\partial}$ in formulas 
\eqref{e:63},  \eqref{e:64},\eqref{e:65} 
operates on the function in such a way: 
$ u\overleftrightarrow{\partial}
v = u (\partial v) - (\partial u) v $.

Lagrangian of $ND_{33}\gamma$ interaction:
\begin{eqnarray}\label{e:64}
& & \mathcal{L}_{ND_{33}\gamma }=\frac{G_{1ND_{33}\gamma }}{4M}\left\{ \overline{%
N}^{\ast \lambda }\left[ \left( g_{\lambda \alpha }\gamma _{\beta
}-g_{\lambda \beta }\gamma _{\alpha }\right) +\frac{1}{2}\gamma _{\lambda
}\sigma _{\alpha \beta }\right] S_{3}^{\dag }N+h.c.\right\} F^{\alpha \beta
}- \nonumber \\
& & -\frac{G_{2ND_{33}\gamma }}{8M^{2}}\left\{ \overline{N}^{\ast \lambda }\left[
i\left( g_{\lambda \beta }\overleftrightarrow{\partial_{\alpha}}-g_{\lambda \alpha }%
\overleftrightarrow{\partial_{\beta}}\right) - \frac{i}{4}\gamma_{\lambda }\left(
\gamma_{\beta }\overleftrightarrow{\partial_{\alpha}} - \gamma _{\alpha }%
\overleftrightarrow{\partial_{\beta}}\right) \right] S_{3}^{\dag }N+h.c.\right\}
F^{\alpha \beta }+ \nonumber \\
& & +\frac{G_{3ND_{33}\gamma }}{4M^{2}}\left\{ \left[ \overline{N}^{\ast \lambda
}i\left( -g_{\lambda \beta }+\frac{1}{4}\gamma _{\lambda }\gamma _{\beta
}\right) S_{3}^{\dag }N+h.c.\right] \partial _{\alpha }F^{\alpha \beta }+ \right. \nonumber\\
& & \left. +\left[ \overline{N}^{\ast \lambda }i\left( g_{\lambda \alpha }-\frac{1}{4}
\gamma _{\lambda }\gamma _{\alpha }\right) S_{3}^{\dag }N+h.c.\right]
\partial _{\beta }F^{\alpha \beta }\right\},
\end{eqnarray}
where $G_{1ND_{33}\gamma},\, G_{2ND_{33}\gamma},\,G_{3ND_{33}\gamma}$
are  isovector  form  factors of  $ND_{33}\gamma$ interaction. 

Lagrangian of $ND_{13}\gamma$ interaction:
\begin{eqnarray}\label{e:65}
& & \mathcal{L}_{ND_{13}\gamma }=\frac{1}{4M}\left\{ \overline{N}^{\ast \lambda }%
\left[ \left( g_{\lambda \alpha }\gamma _{\beta }-g_{\lambda \beta }\gamma
_{\alpha }\right) +\frac{1}{2}\gamma _{\lambda }\sigma _{\alpha \beta }%
\right] \frac{1}{2}\left[ G_{1ND_{13}\gamma }^{S}+G_{1ND_{13}\gamma
}^{V}\tau _{3}\right] N+h.c.\right\} F^{\alpha \beta } - \nonumber \\
& & - \frac{1}{8M^{2}}\left\{ \overline{N}^{\ast \lambda }\left[i \left(
g_{\lambda \beta }\overleftrightarrow{\partial_{\alpha}}-g_{\lambda \alpha }%
\overleftrightarrow{\partial_{\beta}}\right) - \frac{i}{4}\gamma_{\lambda }\left(
\gamma_{\beta }\overleftrightarrow{\partial_{\alpha}}-\gamma _{\alpha }%
\overleftrightarrow{\partial_{\beta}}\right) \right] \times \right. \nonumber \\
& & \left. \times \frac{1}{2}\left[
G_{2ND_{13}\gamma }^{S}+G_{2ND_{13}\gamma }^{V}\tau _{3}\right]
N+h.c.\right\} F^{\alpha \beta }+ \nonumber \\
& & +\frac{1}{4M^{2}}\left[ \overline{N}^{\ast \lambda }i\left( -g_{\lambda
\beta }+\frac{1}{4}\gamma _{\lambda }\gamma _{\beta }\right) \frac{1}{2}%
\left[ G_{3ND_{13}\gamma }^{S}+G_{3ND_{13}\gamma }^{V}\tau _{3}\right] N+h.c.%
\right] \partial _{\alpha }F^{\alpha \beta }+ \nonumber \\
& & +\frac{1}{4M^{2}}\left[ \overline{N}^{\ast \lambda }i\left( g_{\lambda
\alpha }-\frac{1}{4}\gamma _{\lambda }\gamma _{\alpha }\right) \frac{1}{2}%
\left[ G_{3ND_{13}\gamma }^{S}+G_{3ND_{13}\gamma }^{V}\tau _{3}\right] N+h.c.%
\right] \partial _{\beta }F^{\alpha \beta },
\end{eqnarray}
where $G_{{i\,ND_{13}\gamma }}^{S} = G_{{i\,ND_{13}\gamma}}^{p} + \; G_{{i\,ND_{13}\gamma}}^{n},\,
G_{{i\,ND_{13}\gamma }}^{V} = G_{{i\,ND_{13}\gamma }}^{p} - \; G_{{i\,ND_{13}\gamma }}^{n}$
are isoscalar and isovector  form  factors  of $NP_{11}\gamma$ 
interaction. 

Lagrangians  \eqref{e:60} - \eqref{e:65} are  gauge-invariant, 
because   four-potential  $A^{\mu}$  is  also  involved due to 
gauge-invariant tensor $F^{\mu \nu}$.  It is necessary to note 
that the contribution of components, proportional  $\partial_{\mu}F^{\mu \nu}$, 
in lagrangians \eqref{e:63} - \eqref{e:65} turns into zero for 
real photons $(k^2 = 0, \, \epsilon \cdot k = 0)$.  

Propagator  was  used  in  this  work  for  $\frac{1}{2}$ spin 
resonances  $(S_{11},\, S_{31}, \, P_{11})$:
\begin{equation}\label{e:66}
G(P^2; P)=\frac{\widehat{P}+M^{*}}{P^2-{M^{*}}^{2}+
i M^{*} \Gamma(P^2)}.
\end{equation}
Rarita-Schwinger  propagator  was  used for $\frac{3}{2}$ spin 
resonances $(P_{33},\, D_{33},\,D_{13})$:
\begin{equation}\label{e:67}
G^{\mu \nu} (P^2; P)=\frac{\widehat{P}+M^{*}}{P^2-{M^{*}}^{2}+
i M^{*} \Gamma(P^2)} \times \left\{-g^{\mu \nu}
+\frac{\gamma^{\mu} \gamma^{\nu}}{3} +
\frac{2 P^{\mu} P^{\nu}}{3 {M^{*}}^{2}}-
\frac{P^{\mu}\gamma^{\nu}-\gamma^{\mu} P^{\nu}}{3 M^{*}}
\right\}.
\end{equation}
In   expressions   \eqref{e:66},   \eqref{e:67}\,  $M^{*}$   É
$\Gamma(P^2)$ are mass and width of the corresponding resonance 
with four-momentum $P$.

\vspace {2.0cm}
Below there are explicit expressions for Born diagrams 
contributions and diagrams of  six resonances 
$P_{33}(1232),\, P_{11}(1430),\,
 S_{11}(1500),\,  S_{31}(1620),\,   
 D_{33}(1700),\, D_{13}(1505)$
into invariant functions of $f_1 - f_6$  amplitude \eqref{e:17}.

Born diagrams contribution to invariant functions  $f_1 - f_6$ 
of Compton scattering on proton:
\begin{eqnarray}\label{e:101}
& & \left\{ \frac{-2\,M\,t\,{F_{1}^{p}}^2}
{\left( M^2 - s \right) \,\left( M^2 - s - t \right) },
\frac{\left( 2\,M^2 - 2\,s - t \right) \,{F_{1}^{p}}^2}
{\left( -M^2 + s \right) \,
\left( -M^2 + s + t \right) }\right.,\nonumber \\
& & \frac{\left( 2\,F_{1}^{p} - F_{2}^{p} \right) \,F_{2}^{p}}{M},
\frac{\left(- 2\,M^2 + 2\,s + t \right) \,
{\left( F_{1}^{p} - F_{2}^{p} \right) }^2}{\left( M^2 - s \right) \,
\left( M^2 - s - t \right) }, \nonumber \\
& &\frac{ 2\,M^2\,t\,{F_{1}^{p}}^2 +
2\,\left( M^4 - 2\,M^2\,\left( s + t \right)  +
s\,\left( s + t \right)  \right) \,F_{1}^{p}\,F_{2}^{p} -
\left( M^2 - s \right) \,\left( M^2 - s - t \right) \,
{F_{2}^{p}}^2  }{2 M\,\left( M^2 - s \right) \,
\left( M^2 - s - t \right) } + \nonumber \\
& & +\frac{2\,f\,M\,t\,{F^{\pi}}}{\mu \,\left( t -
{\mu }^2 \right) } +
\frac{2\,F^{\eta }\,M\,t\,{f_{\eta }}}
{{{\mu }_{\eta }}\,\left( t - {{{\mu }_{\eta }}}^2 \right) }, \nonumber\\
& &\left. \frac{ -2\,M^2\,t\,{F_{1}^{p}}^2 +
2\,M^2\,t\,F_{1}^{p}\,F_{2}^{p} +
\left( M^2 - s \right) \,\left( M^2 - s - t \right) \,{F_{2}^{p}}^2 \
 }{2 M^2\,\left( M^2 - s \right) \,
\left( M^2 - s - t \right) } \right\}.
\end{eqnarray}

It   is   necessary   to   make  a  change in  case of Compton 
scattering   on   neutron   in   formulae   \eqref{e:101}:
\begin{equation}
F_{1}^{p} \rightarrow F_{1}^{n},\quad F_{2}^{p} \rightarrow F_{2}^{n},
\quad f \rightarrow -f.
\end{equation}

Contribution of $P_{33}$ into invariant functions $f_1 - f_6$ 
of Compton scattering ( is the same for Compton scattering 
on proton and neutron):
\begin{eqnarray}
 & & \left\{ -\frac{1}{576\,M^4\,{D_{s\Delta }}\,{{M_{\Delta }}}^2}
\,\left( 16\,M^2\,\left( M - {M_{\Delta }} \right) \,
 \left( M^4 - M^2\,s + 3\,\left( M^3 - M\,s \right) \,{M_{\Delta }} +
 3\,t\,{{M_{\Delta }}}^2 \right) \,{G_{\Delta N \gamma}^{1}}^2 -
 \right. \right.
\nonumber \\
& & \left. -16\,M\,\left( M^2\,\left( M^2 - s \right) \,s + \left( M^5 - M\,s^2 \right) \,{M_{\Delta }} -
 3\,s\,\left( -M^2 + s + 2\,t \right) \,{{M_{\Delta }}}^2 + \right. \right.
\nonumber \\
& & \left. \left. +6\,M\,\left( -M^2 + s + t \right) \,{{M_{\Delta }}}^3 \right) \,
 G_{\Delta N \gamma }^{1}\,G_{\Delta N \gamma }^{2} +
 \left( M\,s\,\left( M^4 + 2\,M^2\,s - 3\,s^2 \right)  - \right. \right. \nonumber \\
& & \left. \left. -2\,s\,\left( -3\,M^4 + 2\,M^2\,s + s^2 \right) \,{M_{\Delta }} +
 3\,M\,\left( 5\,M^4 - 22\,M^2\,s + s\,\left( 17\,s + 16\,t \right)  \right) \,
 {{M_{\Delta }}}^2 - \right. \right.
\nonumber \\
& & \left. \left. -6\,\left( M^4 - 6\,M^2\,s + s\,\left( 5\,s
+ 8\,t \right)  \right) \,{{M_{\Delta }}}^3 \right)
{G_{\Delta N \gamma }^{2}}^2 \right) + \nonumber\\
& & + \frac{1}
{576\,M^4\,{D_{u\Delta }}\,{{M_{\Delta }}}^2}
\,\left( 16\,M^2\,\left( M - {M_{\Delta }} \right) \,
 \left( M^2\,\left( M^2 - s - t \right)  + 3\,M\,\left( M^2 - s - t \right) \,{M_{\Delta }} -
 3\,t\,{{M_{\Delta }}}^2 \right) \times \right.
\nonumber \\
& & \left. \times {G_{\Delta N \gamma }^{1}}^2 - 16\,M\,
\left( M^2\,\left( 2\,M^4 - 3\,M^2\,\left( s + t \right)  +
 {\left( s + t \right) }^2 \right)  +
 M\,\left( 3\,M^4 - 4\,M^2\,\left( s + t \right)  + \right.
\right. \right. \nonumber \\
& & \left. \left. \left.
+ {\left( s + t \right) }^2 \right) \,
{M_{\Delta }} + 3\,\left( 2\,M^4 + s^2 - t^2 + M^2\,
\left( t - 3\,s \right)  \right) \,
{{M_{\Delta }}}^2 - 6\,\left( M^3 - M\,s \right) \,{{M_{\Delta }}}^3 \right) \,
 G_{\Delta N \gamma }^{1}\,G_{\Delta N \gamma }^{2} + \right. \nonumber \\
& & \left.  +\left( M\,\left( 14\,M^6 - 27\,M^4\,\left( s + t \right)
  + 16\,M^2\,{\left( s + t \right) }^2 - 3\,{\left( s + t \right) }^3 \right)  +
  \left( 20\,M^6 - 34\,M^4\,\left( s + t \right)  + \right. \right. \right. \nonumber \\
& & \left. \left. \left. + 16\,M^2\,{\left( s + t \right) }^2 -
  2\,{\left( s + t \right) }^3 \right) \,{M_{\Delta }} -
  3\,M\,\left( 29\,M^4 + 17\,s^2 + 18\,s\,t + t^2 - 2\,M^2\,\left( 23\,s + 7\,t \right)
  \right) \,{{M_{\Delta }}}^2 + \right. \right. \nonumber \\
& &  \left. \left. +  6\,\left( 9\,M^4 + 5\,s^2 + 2\,s\,t - 3\,t^2 + 2\,M^2\,\left( -7\,s + t \right)  \right) \,
  {{M_{\Delta }}}^3 \right) \,{G_{\Delta N \gamma}^{2}}^2
  \right), \nonumber
\end{eqnarray}
\begin{eqnarray}
& & -\frac{1}{576\,M^4\,{D_{s\Delta }}\,{{M_{\Delta }}}^2} \left(
16\,M^2\,\left( M^2\,\left( M^2 + s \right)  + 4\,M^3\,{M_{\Delta }} +
3\,\left( -3\,M^2 + s + t \right) \,{{M_{\Delta }}}^2 \right) \,
{G_{\Delta N \gamma }^{1}}^2 - \right. \nonumber \\
& & \left. -8\,M\,
\left( M\,s\,\left( 3\,M^2 + s \right)  + 2\,\left( M^4 + 3\,M^2\,s \right) \,{M_{\Delta }} +
3\,M\,\left( 3\,M^2 + s \right) \,{{M_{\Delta }}}^2 + \right. \right.
\nonumber \\
& & \left. \left.
+6\,\left( -5\,M^2 + s + 2\,t \right) \,{{M_{\Delta }}}^3 \right) \,
G_{\Delta N \gamma}^{1}\,G_{\Delta N \gamma}^{2} +
\left( s\,\left( M^4 + 6\,M^2\,s + s^2 \right)  +
8\,M\,s\,\left( M^2 + s \right) \,{M_{\Delta }} +
\right. \right. \nonumber \\
& & \left. \left. + 3\,\left( 5\,M^4 - 34\,M^2\,s + s\,\left( 5\,s + 16\,t \right)  \right) \,{{M_{\Delta }}}^2 +
24\,M\,\left( M^2 +
s \right) \,{{M_{\Delta }}}^3 \right)
\,{G_{\Delta N \gamma}^{2}}^2 \right) + \nonumber \\
& & + \frac{1}{576\,M^4\,{D_{u \Delta }}\,{{M_{\Delta }}}^2} \,
  \left( 16\,\left( M^4\,\left( 3\,M^2 - s - t \right)  + 4\,M^5\,{M_{\Delta }} -
  3\,M^2\,\left( M^2 + s \right) \,{{M_{\Delta }}}^2 \right)
  \,{G_{\Delta N \gamma}^{1}}^2 - \right. \nonumber \\
& & \left.  - 8\,M\,\left( M\,\left( 10\,M^4 - 7\,M^2\,\left( s + t \right)  +
  {\left( s + t \right) }^2 \right)  +
  2\,M^2\,\left( 7\,M^2 - 3\,\left( s + t \right)  \right)
  \,{M_{\Delta }} + \right. \right. \nonumber \\
& & \left. \left. + 3\,M\,\left( 5\,M^2 - s - t \right) \,{{M_{\Delta }}}^2 -
  6\,\left( 3\,M^2 + s - t \right) \,{{M_{\Delta }}}^3 \right)
  \,G_{\Delta N \gamma }^{1}\,
  G_{\Delta N \gamma}^{2} + \right. \nonumber \\
& & \left.  + \left( 34\,M^6 - 37\,M^4\,\left( s + t \right)  +
  12\,M^2\,{\left( s + t \right) }^2 - {\left( s + t \right) }^3 +
  \right. \right. \nonumber \\
& & \left. \left. +  8\,M\,\left( 6\,M^4 - 5\,M^2\,\left( s + t \right)  + {\left( s + t \right) }^2 \right) \,
  {M_{\Delta }} - 3\,\left( 43\,M^4 - 5\,s^2 + 6\,s\,t +
\right. \right. \right. \nonumber \\
& & \left. \left. \left.  + 11\,t^2 -
  2\,M^2\,\left( 7\,s + 23\,t \right)  \right) \,{{M_{\Delta }}}^2 +
  24\,M\,\left( 3\,M^2 - s - t \right) \,{{M_{\Delta }}}^3 \right) \,
  {G_{\Delta N \gamma }^{2}}^2 \right), \nonumber 
\end{eqnarray}
\begin{eqnarray}
& & \frac{1}{576\,M^4\,{D_{s \Delta }}\,{{M_{\Delta }}}^2} \,
   \left( 16\,M^2\,\left( M + {M_{\Delta }} \right) \,
   \left( M^4 - M^2\,s - 3\,\left( M^3 - M\,s \right) \,{M_{\Delta }} +
   3\,t\,{{M_{\Delta }}}^2 \right) \,{G_{\Delta N \gamma}^{1}}^2 -
\right. \nonumber \\
& & \left. - 16\,M\,{\left( M^2 - s \right) }^2\,\left( M - 3\,{M_{\Delta }} \right) \,{M_{\Delta }}\,
   G_{\Delta N \gamma}^{1}\,G_{\Delta N \gamma}^{2} +
   {\left( M^2 - s \right) }^2\,\left( M\,s + 2\,s\,{M_{\Delta }} +
\right. \right. \nonumber \\
& & \left. \left. + 15\,M\,{{M_{\Delta }}}^2 -
   18\,{{M_{\Delta }}}^3 \right) \,{G_{\Delta N \gamma}^{2}}^2 \right)-
\nonumber \\
& & -\frac{1}{576\,M^4\,{D_{u \Delta }}\,{{M_{\Delta }}}^2}
 \,\left( 16\,M^2\,\left( M + {M_{\Delta }} \right) \,
  \left( M^2\,\left( M^2 - s - t \right)  + 3\,M\,\left( -M^2 + s + t \right) \,{M_{\Delta }}-
\right. \right. \nonumber \\
& & \left. \left.
- 3\,t\,{{M_{\Delta }}}^2 \right) \,{G_{\Delta N \gamma}^{1}}^2
+ 16\,M\,{\left( -M^2 + s + t \right) }^2\,\left( M - 3\,{M_{\Delta }} \right) \,{M_{\Delta }}\,
  G_{\Delta N \gamma}^{1}\,G_{\Delta N \gamma}^{2} -
  {\left( -M^2 + s + t \right) }^2\, \times
\right.  \nonumber \\
& & \left. \times \left( M\,\left( 2\,M^2 - s - t \right) \left( 4\,M^2 - 2\,\left( s + t \right)  \right) \,{M_{\Delta }} +
   15\,M\,{{M_{\Delta }}}^2 - 18\,{{M_{\Delta }}}^3 \right) \,
  {G_{\Delta N\gamma}^{2}}^2\right), \nonumber
\end{eqnarray}
\begin{eqnarray} \label{e:75}
& &  \frac{1}{576\,M^4\,{D_{s \Delta }}\,{{M_{\Delta }}}^2}
  \,\left( 16\,M^2\,\left( M^2\,\left( M^2 + s \right)  - 4\,M^3\,{M_{\Delta }} +
  3\,\left( -3\,M^2 + s + t \right) \,{{M_{\Delta }}}^2 \right) \,
  {G_{\Delta N \gamma}^{1}}^2 -
\right.  \nonumber \\
& & \left. -  8\,M\,
  \left( M\,s\,\left( -M^2 + s \right)  + 2\,\left( M^4 - M^2\,s \right) \,{M_{\Delta }} -
  3\,\left( M^3 - M\,s \right) \,{{M_{\Delta }}}^2 -
\right. \right. \nonumber \\
& & \left. \left. - 6\,\left( M^2 - s \right) \,{{M_{\Delta }}}^3 \right) \,
  G_{\Delta N \gamma }^{1}\,
  G_{\Delta N \gamma}^{2} + {\left( M^2 - s \right) }^2\,
  \left( s + 15\,{{M_{\Delta }}}^2 \right) \,{G_{\Delta N \gamma}^{2}}^2 \right)-
\nonumber \\
& & -\frac{1}{576\,M^4\,{D_{u \Delta }}\,{{M_{\Delta }}}^2}
\left( 16\,\left( M^4\,\left( 3\,M^2 - s - t \right)  - 4\,M^5\,{M_{\Delta }} -
  3\,M^2\,\left( M^2 + s \right) \,{{M_{\Delta }}}^2 \right) \,
 {G_{\Delta N \gamma}^{1}}^2 -
 \right. \nonumber \\
& & \left. - 8\,M\,\left( M\,\left( 2\,M^4 - 3\,M^2\,\left( s + t \right)  +
  {\left( s + t \right) }^2 \right)  +
  2\,M^2\,\left( -M^2 + s + t \right) \,{M_{\Delta }} +
\right. \right. \nonumber \\
& & \left. \left. + 3\,M\,\left( M^2 - s - t \right) \,{{M_{\Delta }}}^2 +
  6\,\left( M^2 - s - t \right) \,{{M_{\Delta }}}^3 \right) \,G_{\Delta N \gamma}^{1}\,
  G_{\Delta N \gamma}^{2} +
\right. \nonumber \\
& & \left. + {\left( -M^2 + s + t \right) }^2\,
  \left( 2\,M^2 - s - t + 15\,{{M_{\Delta }}}^2 \right)
\,{G_{\Delta N \gamma}^{2}}^2 \right), \nonumber
\end{eqnarray}
\begin{eqnarray}
& & -\frac{1}{288\,M^4\,{D_{s \Delta }}\,{{M_{\Delta }}}^2} \,
   \left( 8\,M^2\,{M_{\Delta }}\,\left( 2\,M^4 - 2\,M^2\,s + 3\,t\,{{M_{\Delta }}}^2 \right)
   \,{G_{\Delta N \gamma}^{1}}^2 -
   4\,M\,\left( M^2\,\left( M^2 - s \right) \,s +
\right. \right. \nonumber \\
& & \left. \left. + {M_{\Delta }}\,\left( 2\,M\,\left( M^2 - s \right) \,s +
   3\,{M_{\Delta }}\,\left( M^4 - M^2\,s + 2\,s\,t +
   2\,M\,\left( -M^2 + s \right) \,{M_{\Delta }} \right)  \right)  \right) \,
\times  \right. \nonumber \\
& & \left. \times  G_{\Delta N \gamma}^{1}\,G_{\Delta N \gamma}^{2} +
   \left( M^2 - s \right) \,\left( M\,s^2 + s\,\left( M^2 + s \right) \,{M_{\Delta }} -
   9\,M\,s\,{{M_{\Delta }}}^2 +
\right. \right. \nonumber \\
& & \left. \left. + 3\,\left( M^2 + s \right) \,{{M_{\Delta }}}^3 \right) \,
   {G_{\Delta N \gamma}^{2}}^2 \right)- \nonumber \\
& & -\frac{1}{288\,M^4\,{D_{u \Delta }}\,{{M_{\Delta }}}^2}
   \,\left( -8\,M^2\,{M_{\Delta }}\,\left( 2\,M^2\,\left( M^2 - s - t \right)  -
   3\,t\,{{M_{\Delta }}}^2 \right) \,{G_{\Delta N \gamma}^{1}}^2 +
\right. \nonumber \\
& & \left. + 4\,M\,\left( M^2\,\left( M^2 - s - t \right) \,\left( 2\,M^2 - s - t \right)  +
   {M_{\Delta }}\,\left( 2\,M\,\left( M^2 - s - t \right) \,\left( 2\,M^2 - s - t \right)  +
\right. \right. \right. \nonumber \\
& & \left. \left. \left. +   3\,{M_{\Delta }}\,\left( M^4 + 2\,t\,\left( s + t \right)  -
   M^2\,\left( s + 5\,t \right)  + 2\,M\,\left( -M^2 + s + t \right) \,{M_{\Delta }}
   \right)  \right)  \right) \,G_{\Delta N \gamma}^{1}\,G_{\Delta N \gamma}^{2} -
\right. \nonumber \\
& & \left. - \left( M^2 - s - t \right) \,\left( M\,{\left( -2\,M^2 + s + t \right) }^2 +
   \left( 2\,M^2 - s - t \right) \,\left( 3\,M^2 - s - t \right) \,{M_{\Delta }} +
\right. \right. \nonumber \\
& & \left. \left.  + 9\,M\,\left( -2\,M^2 + s + t \right) \,{{M_{\Delta }}}^2 +
   \left( 9\,M^2 - 3\,\left( s + t \right)  \right) \,{{M_{\Delta }}}^3 \right) \,
   {G_{\Delta N \gamma}^{2}}^2 \right), \nonumber 
\end{eqnarray}
\begin{eqnarray}
& & -\frac{1}{576\,M^4\,{D_{s \Delta }}\,{{M_{\Delta }}}^2} \,
      \left( 16\,M^2\,\left( M^4 - M^2\,s +
       3\,\left( -M^2 + s + t \right) \,{{M_{\Delta }}}^2 \right) \,{G_{\Delta N \gamma}^{1}}^2
-  \right. \nonumber \\
& & \left. - 8\,M\,\left( M\,\left( M^2 - s \right) \,s +
       2\,\left( M^4 - M^2\,s \right) \,{M_{\Delta }} +
       3\,\left( M^3 - M\,s \right) \,{{M_{\Delta }}}^2 +
\right. \right. \nonumber \\
& & \left. \left. + 6\,\left( -M^2 + s + t \right) \,{{M_{\Delta }}}^3 \right) \,G_{\Delta N \gamma}^{1}\,
       G_{\Delta N \gamma}^{2} + \left( s\,\left( M^4 - s^2 \right)  +
       4\,M\,\left( M^2 - s \right) \,s\,{M_{\Delta }} -
\right. \right. \nonumber \\
& & \left. \left. - 9\,\left( M^4 - s^2 \right) \,{{M_{\Delta }}}^2 +
       12\,\left( M^3 - M\,s \right) \,{{M_{\Delta }}}^3 \right) \,{G_{\Delta N \gamma}^{2}}^2
    \right) + \nonumber \\
& & + \frac{1}{576\,M^4\,{D_{u \Delta }}\,{{M_{\Delta }}}^2} \,
       \left( 16\,\left( M^4\,\left( M^2 - s - t \right)  -
       3\,\left( M^4 - M^2\,s \right) \,{{M_{\Delta }}}^2 \right) \,{G_{\Delta N \gamma}^{1}}^2
- \right. \nonumber \\
& & \left. - 8\,M\,\left( M\,\left( 2\,M^4 - 3\,M^2\,\left( s + t \right)  +
       {\left( s + t \right) }^2 \right)  + 2\,M^2\,\left( M^2 - s - t \right) \,{M_{\Delta }} +
\right. \right. \nonumber \\
& & \left. \left. + 3\,M\,\left( M^2 - s - t \right) \,{{M_{\Delta }}}^2 -
       6\,\left( M^2 - s \right) \,{{M_{\Delta }}}^3 \right) \,G_{\Delta N \gamma}^{1}\,
       G_{\Delta N \gamma }^{2} +
\right. \nonumber \\
& & \left. + \left( 6\,M^6 - 11\,M^4\,\left( s + t \right)  +
       6\,M^2\,{\left( s + t \right) }^2 - {\left( s + t \right) }^3 +
       4\,M\,\left( 2\,M^4 -
\right. \right.  \right. \nonumber \\
& & \left. \left. \left. - 3\,M^2\,\left( s + t \right)  + {\left( s + t \right) }^2 \right) \,
       {M_{\Delta }} - 9\,\left( 3\,M^4 - 4\,M^2\,\left( s + t \right)  +
       {\left( s + t \right) }^2 \right) \,{{M_{\Delta }}}^2 +
\right.  \right. \nonumber \\
& & \biggl. \left. \left.    + 12\,M\,\left( M^2 - s - t \right) \,{{M_{\Delta }}}^3 \right) \,
       {G_{\Delta N \gamma}^{2}}^2 \right) \biggr\}.
\end{eqnarray}
In formulae \eqref{e:75} $M_{\Delta}$ is $P_{33}$ resonance mass,
$ D_{s \Delta} = s - M_{\Delta}^{2} + i M_{\Delta} \Gamma_{\Delta}(s), \,
D_{u \Delta}=u-M_{\Delta}^{2}+i M_{\Delta}\Gamma_{\Delta}(u)$ 
are denominators of $P_{33}$ resonance propagator  in $s$- and 
$u$-channel correspondingly.

$P_{11}$  contribution  to  invariant functions $f_1 - f_6$ of 
Compton scattering on proton:
\begin{eqnarray} \label{e:104}
& & \left\{ -\frac{\left( M^2 - s \right) \,{{G_{P11p \gamma}}}^2\,
\left( M - {M_{P11}} \right) }{4 M^2\,{D_{sP11}}} +
\frac{\left( M^2 - s - t \right) \,{{G_{P11p \gamma}}}^2\,
\left( M - {M_{P11}} \right) }{4 M^2\,{D_{uP11}}}\right.,
\nonumber \\
& & -\frac{{{G_{P11p \gamma}}}^2\,
\left( M^2 + s - 2\,M\,{M_{P11}} \right) }{4M^2\,
{D_{sP11}}} +
\frac{{{G_{P11p \gamma}}}^2\,
\left( 3\,M^2 - s - t - 2\,M\,{M_{P11}} \right) }{4M^2\,
{D_{uP11}}},
\nonumber \\
& &  \frac{\left( M^2 - s \right) \,{{G_{P11p \gamma}}}^2\,
   \left( M + {M_{P11}} \right) }{4 M^2\,{D_{sP11}}} -
   \frac{\left( M^2 - s - t \right) \,{{G_{P11p \gamma}}}^2\,
   \left( M + {M_{P11}} \right) }{4 M^2\,{D_{uP11}}},
\nonumber \\
& & \frac{{{G_{P11p \gamma}}}^2\,
  \left( M^2 + s + 2\,M\,{M_{P11}} \right) }{4 M^2\,
  {D_{sP11}}} -
   \frac{{{G_{P11p \gamma}}}^2\,
   \left( 3\,M^2 - s - t + 2\,M\,{M_{P11}} \right) }{4 M^2\,
  {D_{uP11}}},
\nonumber \\
& &  \frac{\left( M^2 - s \right) \,{{G_{P11p \gamma}}}^2\,{M_{P11}}}
    {4 M^2\,{D_{sP11}}} - \frac{\left( M^2 - s - t \right) \,
    {{G_{P11p \gamma}}}^2\,{M_{P11}}}{4 M^2\,{D_{uP11}}},
\nonumber \\
& & \left. -\frac{\left( M^2 - s \right) \,{{G_{P11p \gamma}}}^2}
   {4 M^2\,{D_{sP11}}} + \frac{\left( M^2 - s - t \right) \,
   {{G_{P11p \gamma}}}^2}{4 M^2\,{D_{uP11}}} \, \right\}.
\end{eqnarray}
In formulae \eqref{e:104} $M_{P11}$  is $P_{11}$ resonance mass, 
$ D_{sP11} = s - M_{P11}^{2} + i M_{P11} \Gamma_{P11}(s), \,
D_{uP11}=u-M_{P11}^{2}+i M_{P11}\Gamma_{P11}(u)$ are denominators 
of resonance  $P_{11}$ propagator in $s$- and $u$-channel 
correspondingly.
It is necessary to make a change in case  of Compton 
scattering  on   neutron  in formulae \eqref{e:104}:
$G_{P11p \gamma}  \rightarrow  G_{P11n \gamma}$.

$S_{11}$ contribution in invariant functions $f_1 - f_6$ of 
Compton scattering on proton:
\begin{eqnarray} \label{e:105}
& & \left\{ -\frac{\,\left( M^2 - s \right) \,{{G_{S11p \gamma}}}^2\,
   \left( M + {M_{{S11}}} \right) }{4 M^2\,{D_{{sS11}}}} +
   \frac{\left( M^2 - s - t \right) \,{{G_{S11p \gamma}}}^2\,
   \left( M + {M_{{S11}}} \right) }{4 M^2\,{D_{{uS11}}}}\right.,
\nonumber \\
& &  -\frac{{{G_{S11p \gamma}}}^2\,
      \left( M^2 + s + 2\,M\,{M_{{S11}}} \right) }{4 M^2\,{D_{{sS11}}}} +
   \frac{{{G_{S11p \gamma}}}^2\,
      \left( 3\,M^2 - s - t + 2\,M\,{M_{{S11}}} \right) }
  {4 M^2\,{D_{{uS11}}}},
\nonumber \\
& &  \frac{\left( M^2 - s \right) \,{{G_{S11p \gamma}}}^2\,
      \left( M - {M_{{S11}}} \right) }{4 M^2\,{D_{{sS11}}}} -
   \frac{\left( M^2 - s - t \right) \,{{G_{S11p \gamma}}}^2\,
      \left( M - {M_{{S11}}} \right) }{4 M^2\,{D_{{uS11}}}},
\nonumber \\
& &  \frac{{{G_{S11p \gamma}}}^2\,
      \left( M^2 + s - 2\,M\,{M_{{S11}}} \right) }{4 M^2\,{D_{{sS11}}}} -
   \frac{{{G_{S11p \gamma}}}^2\,
      \left( 3\,M^2 - s - t - 2\,M\,{M_{{S11}}} \right) }{4 M^2\,{D_{{uS11}}}},
\nonumber \\
& &  -\frac{\left( M^2 - s \right) \,{{G_{S11p \gamma}}}^2\,{M_{{S11}}}}
    {4 M^2\,{D_{{sS11}}}} + \frac{\left( M^2 - s - t \right) \,
      {{G_{S11p \gamma}}}^2\,{M_{{S11}}}}{4 M^2\,{D_{{uS11}}}},
\nonumber \\
& & \left. -\frac{\left( M^2 - s \right) \,{{G_{S11p \gamma}}}^2}
    {4 M^2\,{D_{{sS11}}}} + \frac{\left( M^2 - s - t \right) \,
      {{G_{S11p \gamma}}}^2}{4 M^2\,{D_{{uS11}}}}\right\}.
\end{eqnarray}
In formulae \eqref{e:105} $M_{S11}$ is $S_{11}$ resonance mass, 
$ D_{sS11} = s - M_{S11}^{2} + i M_{S11} \Gamma_{S11}(s), \,
D_{uS11}=u-M_{S11}^{2}+i M_{S11}\Gamma_{S11}(u)$ are denominators 
of $S_{11}$ resonance propagator in $s$- and $u$-channel correspondingly.
It is necessary to make a change in case of Compton 
scattering on neutron in formulae \eqref{e:105}:
$G_{S11p \gamma} \rightarrow G_{S11n \gamma}$. 

$S_{31}$ contribution in invariant functions $f_1 - f_6$ of 
Compton scattering (is the same for Compton scattering on 
proton and neutron):
\begin{eqnarray}\label{e:106}
& & \left\{ -\frac{2\,\left( M^2 - s \right)
\,{{G_{S31N \gamma}}}^2\,\left( M + {M_{{S31}}} \right) }
{3 M^2\,{D_{{sS31}}}} + \frac{ 2\,
\left( M^2 - s - t \right) \,{{G_{S31N \gamma}}}^2\,
\left( M + {M_{{S31}}} \right) }{3 M^2\,{D_{{uS31}}}}\right.,
\nonumber\\ & &
-\frac{2\,{{G_{S31N \gamma}}}^2\,
\left( M^2 + s + 2\,M\,{M_{{S31}}} \right) }
{3 M^2\,{D_{{sS31}}}} +
\frac{2\,{{G_{S31N \gamma}}}^2
\,\left( 3\,M^2 - s - t + 2\,M\,{M_{{S31}}} \right) }
{3 M^2\,{D_{{uS31}}}},
\nonumber\\ & &
\frac{2\,\left( M^2 - s \right) \,
{{G_{S31N \gamma}}}^2\,\left( M - {M_{{S31}}} \right) }
{3 M^2\,{D_{{sS31}}}} -
\frac{2\,
\left( M^2 - s - t \right) \,{{G_{S31N \gamma}}}^2\,
\left( M - {M_{{S31}}} \right) }{3 M^2\,{D_{{uS31}}}},
\nonumber\\ & &
\frac{2\,{{G_{S31N \gamma}}}^2\,
\left( M^2 + s - 2\,M\,{M_{{S31}}} \right) }
{3 M^2\,{D_{{sS31}}}} -
\frac{2\,{{G_{S31N \gamma}}}^2\,
\left( 3\,M^2 - s - t - 2\,M\,{M_{{S31}}} \right) }
{3 M^2\,{D_{{uS31}}}},
\nonumber\\ & &
-\frac{2\,\left( M^2 - s \right) \,
{{G_{S31N \gamma}}}^2\,{M_{{S31}}}}{3 M^2\,{D_{{sS31}}}} +
\frac{2\,\left( M^2 - s - t \right) \,
{{G_{S31N \gamma}}}^2\,{M_{{S31}}}}{3 M^2\,{D_{{uS31}}}},
\nonumber\\ & &
\left. -\frac{2\,\left( M^2 - s \right) \,
{{G_{S31N \gamma}}}^2}{3 M^2\,{D_{{sS31}}}} +
\frac{2\,\left( M^2 - s - t \right) \,
{{G_{S31N \gamma}}}^2}{3 M^2\,{D_{{uS31}}}} \right\}.
\end{eqnarray}
In formulae \eqref{e:106} $M_{S31} $ is $S_{31}$ resonance mass,
$ D_{sS31} = s - M_{S31}^{2} + i M_{S31} \Gamma_{S31}(s), \,
D_{uS31}=u-M_{S31}^{2}+i M_{S31}\Gamma_{S31}(u)$ are denominators 
of  $S_{31}$  resonance  propagator  in  $s$-  and $u$-channel 
correspondingly.

$D_{33}$  contribution  to invariant functions $f_1 - f_6$  of 
Compton scattering (is the same for Compton scattering on proton 
and neutron):
\begin{eqnarray}
& & \biggl\{ \biggr.  \frac{1}{576\,M^4\,{D_{{sD33}}}\,
    {{M_{{D33}}}}^2}
   \left( 16\,M^2\,\left( M + {M_{{D33}}} \right) \,
   \left( M^4 + 3\,M^2\,s - 4\,s^2 - 3\,\left( M^3 - M\,s
   \right) \,{M_{{D33}}} +
\right. \right. \nonumber \\  & & \left. \left. +
   \left( -4\,M^2 + 4\,s + 3\,t \right) \,{{M_{{D33}}}}^2
   \right) \,{G_{D33 N \gamma}^{1}}^2 -
  8\,M\,\left( s\,\left( -3\,M^4 + 2\,M^2\,s + s^2 \right)  +
\right. \right. \nonumber \\  & & \left. \left. +
  2\,\left( M^5 - M\,s^2 \right) \,{M_{{D33}}} +
   \left( M^4 - 6\,M^2\,s + s\,\left( 5\,s + 12\,t \right)
   \right) \,{{M_{{D33}}}}^2 -
   12\,M\,\left( M^2 - s - t \right) \,
   {{M_{{D33}}}}^3 \right) \times
 \right. \nonumber \\  & & \left. \times
   \,G_{D33 N \gamma}^{1}\,G_{D33 N \gamma}^{2} +
   \left( M\,s\,\left( M^4 + 2\,M^2\,s - 3\,s^2 \right)  +
   2\,s\,\left(2\,M^2\,s + s^2 -3\,M^4 \right) \,
   {M_{{D33}}} +
   3\,M\,\left( 5\,M^4 -
\right. \right. \right. \nonumber \\  & & \left. \left. \left. -
   22\,M^2\,s + s\,\left( 17\,s + 16\,t
   \right)  \right) \,{{M_{{D33}}}}^2 +
   6\,\left( M^4 - 6\,M^2\,s + s\,\left( 5\,s + 8\,t \right)
   \right) \,{{M_{{D33}}}}^3 \right) \,{G_{D33 N \gamma}^{2}}^2
   \right)+ \nonumber \\
& & +\frac{1}{576\,M^4\,{D_{{uD33}}}\,
    {{M_{{D33}}}}^2}
   \left( 16\,M^2\,\left( M + {M_{{D33}}} \right) \,
   \left( 9\,M^4 - 13\,M^2\,\left( s + t \right)  +
   4\,{\left( s + t \right) }^2 + 3\,M\,\left( s - M^2 +
 \right. \right. \right. \nonumber \\ & & \left. \left. \left. +
   t \right)\,{M_{{D33}}} +
   \left(4\,s + t - 4\,M^2 \right) \,{{M_{{D33}}}}^2 \right) \,
   {G_{D33 N \gamma}^{1}}^2 +
   8\,M\,\left( 10\,M^6 - 17\,M^4\,\left( s + t \right)  + 8
   \,M^2\,{\left( s + t \right) }^2 -
\right. \right. \nonumber \\ & & \left. \left. -
   {\left( s + t \right) }^3 -
   2\,M\,\left( 3\,M^4 - 4\,M^2\,\left( s + t \right)  +
   {\left( s + t \right) }^2 \right) \,{M_{{D33}}} +
   \left( 9\,M^4 + 5\,s^2 - 2\,M^2\,\left( 7\,s - 5\,t \right)  -
\right. \right. \right. \nonumber \\ & & \left. \left. \left. -
   2\,s\,t - 7\,t^2 \right) \,{{M_{{D33}}}}^2 +
   12\,\left( M^3 - M\,s \right) \,{{M_{{D33}}}}^3 \right)
   \,G_{D33 N \gamma}^{1}\,G_{D33 N \gamma}^{2} +
   \left( M\,\left( 14\,M^6 - 27\,M^4\,\left( s + t \right)  +
\right. \right. \right. \nonumber \\ & & \left. \left. \left. +
   16\,M^2\,{\left( s + t \right) }^2 - 3\,{\left( s + t \right) }^3
   \right)  +
   2\,\left( -10\,M^6 + 17\,M^4\,\left( s + t \right)  - 8\,M^2\,
   {\left( s + t \right) }^2 + {\left( s + t \right) }^3 \right)\,
   {M_{{D33}}} -
\right. \right. \nonumber \\ & & \left. \left. -
    3\,M\,\left( 29\,M^4 + 17\,s^2 + 18\,s\,t + t^2 -
   2\,M^2\,\left( 23\,s + 7\,t \right)  \right) \,
   {{M_{{D33}}}}^2 - 6\,\left( 9\,M^4 + 5\,s^2 + 2\,s\,t -
\right. \right. \right. \nonumber \\ & & \left. \left. \left. -
   3\,t^2 + 2\,M^2\,\left( -7\,s + t \right)  \right) \,
   {{M_{{D33}}}}^3 \right) \,{G_{D33N \gamma}^{2}}^2 \right),
   \nonumber
\end{eqnarray}
\begin{eqnarray}
& &  \frac{1}{576\,M^4\,{D_{{sD33}}}\,
    {{M_{{D33}}}}^2}
    \left( 16\,M^2\,\left( M^4 + 7\,M^2\,s + 2\,s^2 -
     4\,\left( M^3 - 2\,M\,s \right) \,{M_{{D33}}} +
     \left( -15\,M^2 + s +
\right. \right. \right. \nonumber \\ & & \left. \left. \left. +
     3\,t \right) \,{{M_{{D33}}}}^2 -
     8\,M\,{{M_{{D33}}}}^3 \right) \,{G_{D33 N \gamma}^{1}}^2 -
     16\,M\,\left( -2\,M\,s\,\left( M^2 + s \right)  +
     \left( M^4 + 3\,M^2\,s \right) \,{M_{{D33}}} -
\right. \right. \nonumber \\ & & \left. \left. -
     4\,M^3\,{{M_{{D33}}}}^2 + 3\,\left( -5\,M^2 + s + 2\,t
     \right) \,{{M_{{D33}}}}^3 \right) \,G_{D33 N \gamma}^{1}\,
     G_{D33 N \gamma}^{2} + \left( s\,\left( M^4 + 6\,M^2\,s +
     s^2 \right)  -
\right. \right. \nonumber \\ & & \left. \left. -
     8\,M\,s\,\left( M^2 + s \right) \,
     {M_{{D33}}} +
     3\,\left( 5\,M^4 - 34\,M^2\,s + s\,\left( 5\,s + 16\,t
     \right)  \right) \,{{M_{{D33}}}}^2 -
     24\,M\,\left( M^2 + s \right)\times
\right. \right. \nonumber \\ & & \left. \left.
    \times {{M_{{D33}}}}^3 \right)
     \,{G_{D33 N \gamma}^{2}}^2 \right)+
   \nonumber \\
& &  +\frac{1}{576\,M^4\,{D_{{uD33}}}\,{{M_{{D33}}}}^2}
     \left( 16\,M^2\,\left( 23\,M^4 - 15\,M^2\,
     \left( s + t \right)  + 2\,{\left( s + t \right) }^2 +
     4\,M\,\left( 3\,M^2 - 2\,\left( s + t \right)  \right) \times
\right. \right. \nonumber \\ & & \left. \left. \times
      \,{M_{{D33}}} - \left( 13\,M^2 + s - 2\,t \right) \,{{M_{{D33}}}}^2 -
      8\,M\,{{M_{{D33}}}}^3 \right) \,{G_{D33 N \gamma}^{1}}^2 +
      16\,M\,\left( 2\,M\,\left( 6\,M^4 - 5\,M^2\,
      \left( s + \right.
\right. \right. \right. \nonumber \\ & & \left. \left. \left.
      \left. + t \right)  + {\left( s + t \right) }^2 \right)  +
      M^2\,\left( -7\,M^2 + 3\,\left( s + t \right)  \right)
      \,{M_{{D33}}} + 4\,M^3\,{{M_{{D33}}}}^2 +
      3\,\left( 3\,M^2 + s - t \right)
      \,{{M_{{D33}}}}^3 \right) \times
\right. \nonumber \\ & & \left. \times
      G_{D33 N \gamma}^{1}\,G_{D33 N \gamma}^{2} +
      \left( 34\,M^6 - 37\,M^4\,\left( s + t \right)  +
      12\,M^2\,{\left( s + t \right) }^2 - {\left( s + t \right) }^3
      - 8\,M\,\left( 6\,M^4 -
\right. \right. \right. \nonumber \\ & & \left. \left. \left. -
      5\,M^2\,\left( s + t \right)  + {\left( s
      + t \right) }^2 \right) \,{M_{{D33}}} -
      3\,\left( 43\,M^4 - 5\,s^2 + 6\,s\,t + 11\,t^2 - 2\,M^2\,
      \left( 7\,s + 23\,t \right)  \right) \,{{M_{{D33}}}}^2 -
\right. \right. \nonumber \\ & & \left. \left. -
      24\,M\,\left( 3\,M^2 - s - t \right) \,{{M_{{D33}}}}^3 \right)
      \,{G_{D33 N \gamma}^{2}}^2 \right),\nonumber
\end{eqnarray}
\begin{eqnarray}
& & \frac{1}{576\,M^4\,{D_{{sD33}}}\,
     {{M_{{D33}}}}^2}
     \left( 16\,M^2\,\left( M^5 - M^3\,s + 2\,\left( M^4 - 3\,
     M^2\,s + 2\,s^2 \right) \,{M_{{D33}}} +
     3\,M\,\left( -M^2 + s +
\right. \right. \right. \nonumber \\ & & \left. \left. \left. +
      t \right) \,{{M_{{D33}}}}^2 +
     \left( 4\,M^2 - 4\,s - 3\,t \right) \,{{M_{{D33}}}}^3 \right) \,
     {G_{D33 N \gamma}^{1}}^2 - 8\,M\,{\left( M^2 - s \right)}^2\,
     \left( -s + 2\,M\,{M_{{D33}}} +
\right. \right. \nonumber \\ & & \left. \left. +
     7\,{{M_{{D33}}}}^2 \right)
     \,G_{D33 N \gamma}^{1}\,G_{D33 N \gamma}^{2} +
     {\left( M^2 - s \right) }^2\,\left( M\,s - 2\,s\,
     {M_{{D33}}} + 15\,M\,{{M_{{D33}}}}^2 + 18\,{{M_{{D33}}}}^3
     \right) \times
\right. \nonumber \\ & & \left. \times
     {G_{D33 N \gamma}^{2}}^2 \right) -
\nonumber \\
& &  -\frac{1}{576\,M^4\,{D_{{uD33}}}\,{{M_{{D33}}}}^2}
     \left( 16\,M^2\,\left( M^3\,\left( M^2 - s - t \right)  -
     2\,\left( 3\,M^4 - 5\,M^2\,\left( s + t \right)  +
     2\,{\left( s + t \right) }^2 \right) \,{M_{{D33}}} -
\right. \right. \nonumber \\ & & \left. \left. -
     3\,\left( M^3 - M\,s \right) \,{{M_{{D33}}}}^2 +
     \left( 4\,M^2 - 4\,s - t \right) \,{{M_{{D33}}}}^3 \right) \,
     {G_{D33 N \gamma}^{1}}^2 - 8\,M\,{\left( -M^2 + s + t \right) }^2\,
     \left( 2\,M^2 - s -
\right. \right. \nonumber \\ & & \left. \left. -
     t - 2\,M\,{M_{{D33}}} -
     7\,{{M_{{D33}}}}^2 \right) \,G_{D33 N \gamma}^{1}\,
     G_{D33 N \gamma}^{2} -
     {\left( -M^2 + s + t \right) }^2\,\left( M\,\left( 2\,M^2
     - s - t \right)  + 2\,\left( -2\,M^2 +
\right. \right. \right. \nonumber \\ & & \left. \left. \left. +
     s + t \right) \,
     {M_{{D33}}} +
     15\,M\,{{M_{{D33}}}}^2 + 18\,{{M_{{D33}}}}^3 \right) \,
     {G_{D33 N \gamma}^{2}}^2 \right), \nonumber
\end{eqnarray}
\begin{eqnarray}
& & \frac{1}{576\,M^4\,{D_{{sD33}}}\,{{M_{{D33}}}}^2}
    \left( 16\,M^2\,\left( M^4 - M^2\,s + 2\,s^2 + 4\,
    \left( M^3 - 2\,M\,s \right) \,{M_{{D33}}} +
    \left( -7\,M^2 + s + 3\,t \right) \times
\right. \right. \nonumber \\ & & \left. \left. \times
    {{M_{{D33}}}}^2 +
    8\,M\,{{M_{{D33}}}}^3 \right) \,{G_{D33 N \gamma}^{1}}^2 -
    16\,M\,\left( M^2 - s \right) \,{M_{{D33}}}\,
    \left( M^2 + 2\,M\,{M_{{D33}}} - 3\,{{M_{{D33}}}}^2 \right) \times
\right. \nonumber \\ & & \left. \times
    G_{D33 N \gamma}^{1}\,G_{D33 N \gamma}^{2} + {\left( M^2 -
    s \right) }^2\,\left( s + 15\,{{M_{{D33}}}}^2 \right) \,
    {G_{D33 N \gamma}^{2}}^2 \right)-
\nonumber \\ & & -
     \frac{1}{576\,M^4\,{D_{{uD33}}}\,{{M_{{D33}}}}^2}
     \left( 16\,M^2\,\left( 7\,M^4 - 7\,M^2\,\left( s + t \right)
     + 2\,{\left( s + t \right) }^2 -
     4\,M\,\left( 3\,M^2 - 2\,\left( s + t \right)  \right) \times
\right. \right. \nonumber \\ & & \left. \left. \times {M_{{D33}}} -
     \left( 5\,M^2 + s - 2\,t \right) \,{{M_{{D33}}}}^2 +
     8\,M\,{{M_{{D33}}}}^3 \right) \,{G_{D33 N \gamma}^{1}}^2 +
     16\,M\,\left( M^2 - s - t \right) \,{M_{{D33}}}\,
     \left( M^2 +
\right. \right. \nonumber \\ & & \left. \left. + 2\,M\,{M_{{D33}}} -
     3\,{{M_{{D33}}}}^2 \right) \,
     G_{D33 N \gamma}^{1}\,G_{D33 N \gamma}^{2} + 
     {\left(s + t - M^2\right) }^2\,
     \left( 2\,M^2 - s - t + 15\,{{M_{{D33}}}}^2 \right) \,
     {G_{D33 N \gamma}^{2}}^2 \right), \nonumber 
\end{eqnarray}
\begin{eqnarray}
& & \frac{1}{288\,M^4\,{D_{{sD33}}}\,{{M_{{D33}}}}^2}
    \left( 8\,M^2\,\left( 2\,M\,s\,\left( s - M^2 \right)  +
    2\,\left( M^4 - 3\,M^2\,s + 2\,s^2 \right) \,{M_{{D33}}} +
    2\,\left( M^3 - M\,s \right) \times
\right. \right. \nonumber \\ & & \left. \left. \times
    {{M_{{D33}}}}^2 +
    \left( 4\,M^2 - 4\,s + 3\,t \right) \,{{M_{{D33}}}}^3
    \right) \,{G_{D33 N \gamma}^{1}}^2 -
    4\,M\,\left( s\,\left( M^4 - s^2 \right)  + 2\,M\,s\,
    \left( -M^2 + s \right) \,{M_{{D33}}} +
\right. \right. \nonumber \\ & & \left. \left. +
    \left( 3\,M^4 - 4\,M^2\,s + s\,\left( s + 6\,t \right)
    \right) \,{{M_{{D33}}}}^2 +
    6\,\left( M^3 - M\,s \right) \,{{M_{{D33}}}}^3 \right)
    \,G_{D33 N \gamma}^{1}\,G_{D33 N \gamma}^{2} +
    \left( M\,s^2\,\left( s -
\right. \right. \right. \nonumber \\ & & \left. \left. \left. -
    M^2 \right)  + s\,
    \left( M^4 - s^2 \right) \,{M_{{D33}}} +
    9\,M\,\left( M^2 - s \right) \,s\,{{M_{{D33}}}}^2 + 3\,
    \left( M^4 - s^2 \right) \,{{M_{{D33}}}}^3 \right) \,
    {G_{D33 N \gamma}^{2}}^2 \right) +
\nonumber \\ & & +
   \frac{1}{288\,M^4\,{D_{{uD33}}}\,{{M_{{D33}}}}^2}
   \left( 8\,M^2\,\left( 2\,M\,\left( 2\,M^4 - 3\,M^2\,
   \left( s + t \right)  + {\left( s + t \right) }^2 \right)  +
   2\,\left( 3\,M^4 - 5\,M^2\,\left( s + t \right)  +
\right. \right. \right. \nonumber \\ & & \left. \left. \left. +
   2\,{\left( s + t \right) }^2 \right) \,{M_{{D33}}} +
   2\,M\,\left( s + t - M^2 \right) \,{{M_{{D33}}}}^2 +
   \left( -4\,M^2 + 4\,s + 7\,t \right) \,{{M_{{D33}}}}^3 \right) \,
   {G_{D33 N \gamma}^{1}}^2 + 4\,M \times
\right. \nonumber \\ & & \left. \times
   \left( 6\,M^6 - 11\,M^4\,
   \left( s + t \right)  + 6\,M^2\,{\left( s + t \right) }^2 -
   {\left( s + t \right) }^3 - 2\,M\,\left( 2\,M^4 - 3\,M^2\,
   \left( s + t \right)  + {\left( s + t \right) }^2 \right) \times
\right. \right. \nonumber \\ & & \left. \left. \times
   {M_{{D33}}} + \left( M^4 - s^2 - 12\,M^2\,t + 4\,s\,t +
   5\,t^2 \right) \,{{M_{{D33}}}}^2 +
   6\,M\,\left( M^2 - s - t \right) \,{{M_{{D33}}}}^3 \right)
   \,G_{D33 N \gamma}^{1} \times 
\right. \nonumber \\ & & \left. \times G_{D33 N \gamma}^{2} +
   \left( M\,\left( M^2 - s - t \right) \,{\left(  s +
   t - 2\,M^2\right) }^2 + \left( -6\,M^6 + 11\,M^4\,\left( s + t
   \right)  - 6\,M^2\,{\left( s + t \right) }^2 + 
\right.
\right. \right. \nonumber \\ & & \left. \left. \left. 
   + \left( s + t \right)^3 \right)
   {M_{{D33}}} - 9\,M\,\left( 2\,M^4 - 3\,M^2\,\left( s +
   t \right)  + {\left( s + t \right) }^2 \right) \,
   {{M_{{D33}}}}^2 - 3\,\left( 3\,M^4 - 4\,M^2\,\left( s +
   t \right)  + \right. \right. \right. \nonumber \\ & & \left. \left. \left.
+ \left( s + t \right)^2 \right)
{{M_{{D33}}}}^3 \right) \,{G_{D33 N \gamma}^{2}}^2 \right),
\nonumber
\end{eqnarray}
\begin{eqnarray}\label{e:107}
& & \frac{1}{576\,M^4\,{D_{{sD33}}}\,{{M_{{D33}}}}^2}
    \left( 16\,M^2\,\left( M^4 + M^2\,s - 2\,s^2 + \left(
    -5\,M^2 + 5\,s + 3\,t \right) \,{{M_{{D33}}}}^2 \right) \,
    {G_{D33 N \gamma}^{1}}^2  -
\right. \nonumber \\ & & \left. -
    16\,M\,\left( M\,s\,\left(
    -M^2 + s \right)  + \left( M^4 - M^2\,s \right) \,
    {M_{{D33}}} +
    \left( -M^3 + M\,s \right) \,{{M_{{D33}}}}^2 + 3\,
    \left( -M^2 + s + t \right) \times
\right. \right. \nonumber \\ & & \left. \left. \times
    {{M_{{D33}}}}^3 \right) \,
    G_{D33 N \gamma}^{1}\,G_{D33 N \gamma}^{2} + \left( s\,
    \left( M^4 - s^2 \right)  +
    4\,M\,s\,\left( -M^2 + s \right) \,{M_{{D33}}} - 9\,
    \left( M^4 - s^2 \right) \,{{M_{{D33}}}}^2 -
\right. \right. \nonumber \\ & & \left. \left. -
    12\,\left( M^3 - M\,s \right) \,{{M_{{D33}}}}^3 \right)
    \,{G_{D33 N \gamma}^{2}}^2 \right) -
\nonumber \\ & &
  -\frac{1}{576\,M^4\,{D_{{uD33}}}\,{{M_{{D33}}}}^2}
   \left( 16\,M^2\,\left( 5\,M^4 - 7\,M^2\,\left( s + t \right)
   + 2\,{\left( s + t \right) }^2 +
   \left( -5\,M^2 + 5\,s + 2\,t \right) \,{{M_{{D33}}}}^2 \right)\times
\right. \nonumber \\ & & \left. \times
   {G_{D33 N \gamma}^{1}}^2 +
   16\,M\,\left( M\,\left( 2\,M^4 - 3\,M^2\,\left( s + t
   \right)  + {\left( s + t \right) }^2 \right)  +
   M^2\,\left( -M^2 + s + t \right) \,{M_{{D33}}} +
\right. \right. \nonumber \\ & & \left. \left. +
   M\,\left(M^2 - s - t \right) \,{{M_{{D33}}}}^2 +
   3\,\left( M^2 - s \right) \,{{M_{{D33}}}}^3 \right) \,
   G_{D33 N \gamma}^{1}\,G_{D33 N \gamma}^{2} +
   \left( 6\,M^6 - 11\,M^4\,\left( s + t \right)  +
\right. \right. \nonumber \\ & & \left. \left. +
   6\,M^2\,{\left( s + t \right) }^2 - {\left( s + t \right) }^3 -
   4\,M\,\left( 2\,M^4 - 3\,M^2\,\left( s + t \right)  +
   {\left( s + t \right) }^2 \right) \,{M_{{D33}}} -
   9\,\left( 3\,M^4 -
\right. \right. \right. \nonumber \\ & & \biggl. \left. \left. \left. -
   4\,M^2\,\left( s + t \right)  +
   {\left( s + t \right) }^2 \right) \,{{M_{{D33}}}}^2 -
   12\,M\,\left( M^2 - s - t \right) \,{{M_{{D33}}}}^3 \right)
   \,{G_{D33 N \gamma}^{2}}^2 \right) \biggr\}.
\end{eqnarray}
In formulae \eqref{e:107} $M_{D33}$  is $D_{33}$ resonance mass, 
$ D_{sD33} = s - M_{D33}^{2} + i M_{D33} \Gamma_{D33}(s), \,
  D_{uD33} = u - M_{D33}^{2} + i M_{D33} \Gamma_{D33}(u)$
are denominators of  $M_{D33}$ resonance propagator in $s$- 
and $u$-channel correspondingly.
Since resonances $D_{33}$ and $D_{13}$ have the same spin and 
parity  $\frac{3}{2}^{-}$ and 
differ only by isospins and masses, $D_{13}$ contribution  to 
invariant functions $f_1 - f_6$ in case of Compton scattering 
on proton may be gained from the corresponding formulae 
\eqref{e:107} for $D_{33}$  by replacement:
\begin{eqnarray}\label{e:108}
& & D_{{sD33}} \rightarrow D_{{sD13}}, \quad
D_{{uD33}} \rightarrow D_{{uD13}}, \quad
M_{D33} \rightarrow M_{D13}, \nonumber \\
& &G_{D33 N \gamma}^{1} \rightarrow \sqrt{\frac{3}{2}}
G_{D13 p \gamma}^{1}, \quad
G_{D33 N \gamma}^{2} \rightarrow \sqrt{\frac{3}{2}}
G_{D13 p \gamma}^{2},
\end{eqnarray}
The second line in $\eqref{e:108}$ must be replaced by the 
following one for Compton scattering on neutron:
\begin{eqnarray}
& &G_{D33 N \gamma}^{1} \rightarrow \sqrt{\frac{3}{2}}
G_{D13 n \gamma}^{1}, \quad
G_{D33 N \gamma}^{2} \rightarrow \sqrt{\frac{3}{2}}
G_{D13 n \gamma}^{2}.
\end{eqnarray}

\end{document}